\documentclass[12pt]{article}
\usepackage{fullpage,citesort,epsfig,psfrag,graphics,amsbsy,amssymb}
\usepackage{caption}
\newcommand{\beq}{\begin{equation}}
\newcommand{\eeq}{\end{equation}}
\newcommand{\bea}{\begin{eqnarray}}
\newcommand{\eea}{\end{eqnarray}}
\newcommand{\bfs}{\boldsymbol}

\newcommand{\be}{\begin{equation}}
\newcommand{\ee}{\end{equation}}
\newcommand{\bq}{\begin{eqnarray}}
\newcommand{\eq}{\end{eqnarray}}

\def\math{\mathsurround=0pt }
\def\leftrightarrowfill{$\math \mathord\leftarrow \mkern-6mu
 \cleaders\hbox{$\mkern-2mu \mathord- \mkern-2mu$}\hfill
 \mkern-6mu \mathord\rightarrow$}
\def\overleftrightarrow#1{\vbox{\ialign{##\crcr
     \leftrightarrowfill\crcr\noalign{\kern-1pt\nointerlineskip}
     $\hfil\displaystyle{#1}\hfil$\crcr}}}

\newcommand{\VEV}[1]{\langle#1\rangle}

\let\l=\lambda

 \def\bd{\begin{document}} \def\ed{\end{document}}
\def\ds{\documentstyle} \let\fr=\frac \let\bl=\bigl \let\br=\bigr
\let\Br=\Bigr \let\Bl=\Bigl
\let\bm=\bibitem
\let\na=\nabla
\let\pa=\partial \let\ov=\overline
\def\ft#1#2{{\textstyle{{\scriptstyle #1}\over {\scriptstyle #2}}}}
\def\fft#1#2{{#1 \over #2}}
\def\vp{\varphi}
\def\sst#1{{\scriptscriptstyle #1}}
\def\oneone{\rlap 1\mkern4mu{\rm l}}
\def\td{\tilde}
\def\wtd{\widetilde}
\def\dalemb#1#2{{\vbox{\hrule height .#2pt
        \hbox{\vrule width.#2pt height#1pt \kern#1pt
                \vrule width.#2pt}
        \hrule height.#2pt}}}
\def\square{\mathord{\dalemb{6.8}{7}\hbox{\hskip1pt}}}
\def\wtd{\widetilde}
\def\R{\rlap{\rm I}\mkern3mu{\rm R}}
\def\im{{\rm i}}
\def\tilg{\tilde{g}}
\def\tilF{\tilde{F}}
\def\tilA{\tilde{A}}
\def\varf{\varphi}
\def\tilf{\tilde{\phi}}
\def\tilh{\tilde{h}}
\def\rme{{\rm e}}
\def\ep{\epsilon}
\def\0{{(0)}}
\def\9{{(9)}}
\def\8{{(8)}}
\def\7{{(7)}}
\def\6{{(6)}}
\def\5{{(5)}}
\def\4{{(4)}}
\def\3{{(3)}}
\def\2{{(2)}}
\def\1{{(1)}}
\newcommand{\trace}{{\rm Tr}}
\newcommand{\ub}{\overline{U}}
\newcommand{\vb}{\overline{V}}
\newcommand{\uh}{\widehat{U}}
\newcommand{\vh}{\widehat{V}}
\newcommand{\ubh}{\overline{\widehat{U}}}
\newcommand{\vbh}{\overline{\widehat{V}}}
\newcommand{\lb}{\bar{\l}}
\newcommand{\Fb}{\overline{F}}
\newcommand{\Fh}{\widehat{F}}
\newcommand{\Fbh}{\overline{\widehat{F}}}
\newcommand{\Ab}{\overline{A}}
\newcommand{\Ah}{\widehat{A}}
\newcommand{\Abh}{\overline{\widehat{A}}}
\newcommand{\Gb}{\overline{G}}
\newcommand{\Gh}{\widehat{G}}
\newcommand{\Gbh}{\overline{\widehat{G}}}
\newcommand{\Pb}{\overline{P}}
\newcommand{\Ph}{\widehat{P}}
\newcommand{\Pbh}{\overline{\widehat{P}}}
\newcommand{\Qb}{\overline{Q}}
\newcommand{\Qh}{\widehat{Q}}
\newcommand{\Qbh}{\overline{\widehat{Q}}}
\newcommand{\Bb}{\overline{B}}
\newcommand{\Bh}{\widehat{B}}
\newcommand{\Bbh}{\overline{\widehat{B}}}
\newcommand{\fhns}{\hat{F}^{\rm (NS)}}
\newcommand{\fhrr}{\hat{F}^{\rm (RR)}}
\newcommand{\ahns}{\hat{A}^{\rm (NS)}}
\newcommand{\ahrr}{\hat{A}^{\rm (RR)}}
\newcommand{\hhrr}{\hat{H}^{\rm (RR)}}
\newcommand{\hchi}{\hat{\chi}}
\newcommand{\hphi}{\hat{\phi}}
\newcommand{\htau}{\hat{\tau}}
\newcommand{\cG}{{\cal G}}
\newcommand{\cGb}{\overline{{\cal G}}}
\newcommand{\cH}{{\cal H}}
\newcommand{\cP}{{\cal P}}
\newcommand{\cPb}{\overline{{\cal P}}}
\newcommand{\cQ}{{\cal Q}}
\newcommand{\cQb}{\overline{{\cal Q}}}
\newcommand{\cM}{{\cal M}}
\newcommand{\cN}{{\cal N}}
\newcommand{\cO}{{\cal O}}
\newcommand{\cD}{{\cal D}}
\newcommand{\cL}{{\cal L}}

\newcommand{\vpp}{\mbox{$\langle{\scriptstyle++}\rangle$}}
\newcommand{\vmp}{\mbox{$\langle{\scriptstyle-+}\rangle$}}
\newcommand{\vppp}{\mbox{$\langle{\scriptstyle+++}\rangle$}}
\newcommand{\vmpp}{\mbox{$\langle{\scriptstyle-++}\rangle$}}
\newcommand{\vpmp}{\mbox{$\langle{\scriptstyle+-+}\rangle$}}

\begin{document}
\setlength{\captionmargin}{36pt}
\begin{titlepage}
\phantom{.}
\vskip 2.5cm
\begin{center}
\begin{large}
{\bf Closed String Self-energy on the Lightcone Worldsheet Lattice}
\footnote{Supported in part by the Department
of Energy under Grant No. DE-FG02-97ER-41029.}
\end{large}

\vskip 2cm
{\large
Georgios Papathanasiou\footnote{E-mail  address: {\tt georgios@ufl.edu}} and Charles B. Thorn\footnote{E-mail  address: {\tt thorn@phys.ufl.edu}}
}
\vskip0.20cm
{\it Institute for Fundamental Theory\\
Department of Physics, University of Florida,
Gainesville FL 32611}

(\today)

\vskip 1.0cm
\end{center}

\begin{abstract}
\noindent We study the one loop correction to the closed
bosonic string propagator, including the possibile presence of D-branes,
by discretizing the light cone worldsheet
on an $M\times N$ rectangular lattice, with $M\propto P^+$ and
$N+1\propto ix^+$.
The integrals over the moduli then become sums which we evaluate numerically.
The main purpose of this study is to assess the reliability of
the worldsheet lattice as a regulator of the divergences in
string perturbation theory. There are two natural geometrical counterterms
for the lightcone worldsheet, one proportional to the area of
the worldsheet and the other proportional to the length of
worldsheet boundaries, tracing the ends of open strings.
We show that the divergences in the closed string self-energy can be cancelled
by the area counterterm and a renormalization of the Regge slope
parameter. The residual finite part is compatible with Lorentz invariance,
provided a novel regularization, natural to the lightcone worldsheet
lattice and described in this article, is employed.

\end{abstract}
\vfill
\end{titlepage}
\section{Introduction}
String theory has long been known to be a generalization of
gauge theory due to
the presence of a massless spin one state in the open string
spectrum. Since all of the massive states of the theory have
masses proportional to $\sqrt{2\pi T_0}=1/\sqrt{\alpha^\prime}$, the
string theory goes over to the gauge theory in the infinite
tension limit $\alpha^\prime\to0$. This, together with the
fact that closed strings have a massless spin two state, has inspired
the discovery of many deep connections between string theory
on the one hand and gauge theory
coupled to gravity on the other.
The AdS/CFT correspondence \cite{maldacenasole}
which asserts the
equivalence of ${\cal N}=4$ supersymmetric gauge theory to
type IIB superstring theory on an AdS$_5\times$S$^5$ space time manifold
is one of the most spectacular of these. Although
motivated by the physical properties of open string theory with
$\alpha^\prime>0$, the final conclusion is reached by taking
$\alpha^\prime\to0$. The finiteness (conformal invariance) of the ${\cal N}=4$
theory plays a key role in justifying the $\alpha^\prime\to0$ limit.

The corresponding hypothesis for an asymptotically free gauge theory
like the gluonic sector of QCD is more obscure. However,  there
is little doubt that string theory can offer important
insights into some aspects of QCD.
In this article we launch a critical analysis of
the possibility that 't Hooft's
$N\to\infty$ limit \cite{thooftlargen} of QCD might
be usefully analyzed by replacing it with the
sum of open string planar diagrams, keeping $\alpha^\prime>0$.
There are several reasons to hope this helps. First, the organization of
multiloop string diagrams is dramatically simpler than the corresponding
gauge theory diagrams: there is only one planar open string diagram
at each loop order, whereas the number of planar gauge theory diagrams
grows exponentially with order. Secondly, the 0 loop open string
planar diagrams describe the evolution of a ``bare'' worldsheet
which becomes ``dressed'' with the inclusion of planar loops.
This provides a very natural setting for the description of
a confining flux tube which may survive the limit $\alpha^\prime\to0$. Finally
there is the long held expectation that the ultraviolet behavior
of gauge theory diagrams will be mitigated by the ``stringiness''
associated with finite $\alpha^\prime$, making the latter better
defined.

We focus on this last point in this article. It is
not so much the ultraviolet divergences themselves that concern us
here--after all those can be absorbed in coupling renormalization
in gauge theories. Rather, it is the extreme care that must
be taken in gauge theory to preserve gauge and Lorentz
invariance in the finite part that remains after renormalization.
Order by order in perturbation theory, this can be
accomplished by employing a suitable regularization, the most popular of which
is dimensional regularization. But as soon as one aims to
extract the consequences of summing all the planar diagrams, especially
if one must rely on numerical methods, the soundness
of dimensional regularization becomes somewhat questionable. For
example the powerful conclusions derived from lattice gauge theory
would be much less convincing if they relied on an unphysical
regularization such as ``analytic
continuation'' of the dimension of spacetime. It is
desirable to have a digitization scheme which can be relied on to
give the correct physical results without such an artifice.
Thirty five years ago Giles and one of us \cite{gilest} (GT) proposed
a digitization of the sum of planar open string diagrams based
on  lightcone quantization \cite{goddardrt,goddardgrt}
in its path history formulation \cite{mandelstamlc}. In
the present article we set out to assess the reliability of this
specific lattice model for perturbative calculations.

\begin{figure}[ht]
\begin{center}
\includegraphics[width=4in]{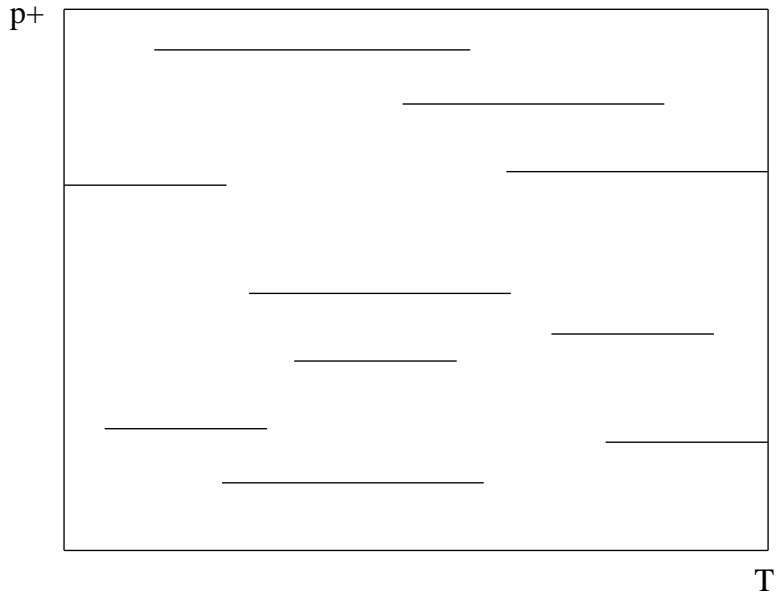}
\caption{Typical open string planar diagram on the lightcone worldsheet. This one
is a seven loop 5 string function}
\label{multiplanar}
\end{center}
\end{figure}

Let us briefly review the GT proposal, in order to set the stage
for the rest of
the paper. In lightcone quantization
of the bosonic string \cite{goddardgrt}, one takes
$x^+=(x^0+x^1)/\sqrt{2}$ as the quantum evolution parameter and labels points
on a string by a parameter $\sigma$ defined so that
$P^+=(P^0+P^1)/\sqrt{2}$
is uniformly distributed on the string. In effect these two choices
eliminate $x^+$ and $x^-$ as dynamical variables, leaving only the $D-2$
transverse coordinates ${\bfs x}(\sigma)$ as quantum operators.
Mandelstam worked out the path history form of this quantization
\cite{mandelstamlc} using imaginary time $\tau\equiv ix^+$.
Then the propagator for a free string
is simply the path integral over the ${\bfs x}(\sigma,\tau)$ where
$0\leq\sigma\leq P^+$ and  $0\leq\tau\leq T$ parameterize a
rectangular region of dimensions $P^+\times T$. The path integrand
is simply $e^{-S}$, with $S$ the lightcone Euclidean action
\bea
S={1\over2}\int_0^T d\tau\int_0^{P^+}d\sigma(\dot{\bfs x}^2+T_0^2{\bfs x}^{\prime2}).
\eea
In this language a general open string planar diagram is calculated by
integrating this same integrand over a worldsheet with several slits
of variable length and location as depicted in Fig~\ref{multiplanar}.

\begin{figure}[ht]
\begin{center}
 \includegraphics[width=4in]{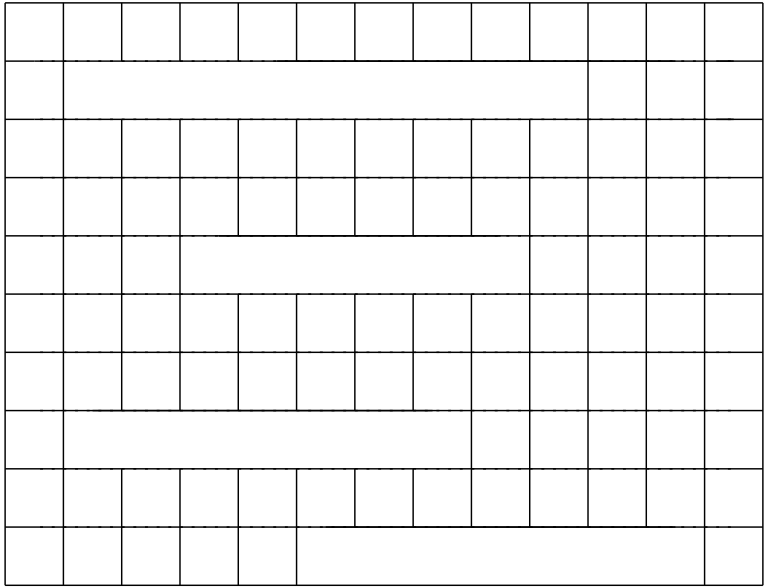}
\caption{Multiloop lattice worldsheet: each loop is a row of missing
links.}
\label{multilooplattice}
\end{center}
\end{figure}

The GT proposal is simply to digitize Mandelstam's interacting string diagrams
by defining a rectangular $M\times N$ grid with $T=(N+1)a$ and $P^+=MaT_0$.
Then the integration variables ${\bfs x}(\sigma,\tau)\to {\bfs x}_{i}^{\ j}$
and the lattice action is simply
\bea
S&\to&{T_0\over2}\sum_{ij}\left[(x_i^{\ j+1}-x_i^{\ j})^2
+(x_{i+1}^{\ j}-x_i^{\ j})^2\right]
\eea
A quick look at the lattice corresponding to a multiloop open string
diagram Fig.~\ref{multilooplattice}, shows that a slit is nothing but a row of missing spatial
bonds (links), and summing over all planar diagrams is simply summing
over all patterns of missing spatial bonds.
One can easily incorporate the sum over missing bond patterns into the
lattice sum over histories by introducing an Ising-like variable
$S_{i}^{\ j}=0,1$ on each spatial link. Then the worldsheet action
describing the sum of all open string planar diagrams is simply
\bea
S_{\rm Planar}&=&{T_0\over2}\sum_{ij}\left[(x_i^{\ j+1}-x_i^{\ j})^2
+S_i^{\ j}(x_{i+1}^{\ j}-x_i^{\ j})^2\right]\nonumber\\
&&\qquad\qquad-\sum_{ij}\left[S_i^{\ j}(1-S_i^{\ j+1})
+S_i^{\ j+1}(1-S_i^{\ j})\right]\ln g
\eea
The purpose of the $\ln g$ term is to insert a factor of $g$ at the
beginning and end of each row of missing bonds.  Then, in addition to
integrating each $x_{i}^{\ j}$ from
$-\infty$ to $+\infty$, one sums each $S_{i}^{\ j}$
over the values 0 and 1 to obtain the sum over all planar diagrams.

If no further adjustments were necessary, we could study this system
numerically, e.g. through Monte Carlo simulation. By analyzing the
large $N$ behavior of the path integral, one could read off spectral
information by identifying exponential behaviors $e^{-aE_\lambda(M)N}$,
where $E_\lambda(M)$ are the eigenvalues of $P^-$. On general grounds
we should expect the large $M$ behavior
\bea
E_\lambda(M)&\sim&\alpha M+\beta +{\gamma_\lambda\over M}+\cdots
\eea
The $\alpha M$ term is a bulk worldsheet effect and the $\beta$ term
is associated with boundaries. The coefficients $\alpha,\beta$
depend on the details of the lattice model and violate the
requirement that $2P^-P^+-{\bfs p}^2=m_\lambda^2$ is a Lorentz
invariant. Fortunately there are two geometrical counterterms
naturally associated with the worldsheet path integral:
one proportional to the area of the worldsheet and another
proportional to the length of worldsheet boundaries.
Thus the Lorentz violating terms noted above can always be cancelled.
As noted in \cite{gilest} the area term is dynamically inconsequential
for the sum of diagrams because the slits representing loops have zero
area, so that the area term is identical for all diagrams contributing
to the same process. The boundary term depends on the number and
lengths of the slits. It can be thought of as an energy cost assigned
to the disappearance of a bond. In the Ising spin description of the
sum over loops it is represented by a term $(D-2)B\sum_{ij}(1-S_i^{\ j})$
added to $S$. The boundary term is absent for closed strings, but
is required for a Lorentz invariant {\it free} open string spectrum.
This is because the Gaussian lattice worldsheet path integral implies
an open string zero point energy
\bea
aP^-(M)&=&(D-2)\sum_{m=1}^{M-1}\sinh^{-1}\sin{m\pi\over 2M}+(D-2)B\nonumber\\
&\sim&(D-2)\left[{2G\over\pi}M+B-{1\over2}\sinh^{-1}1-{\pi\over24M}\right]
\eea
as $M\to\infty$, where $G$ is Catalan's constant. Cancelling the second term requires
$B\to B_0=(1/2)\sinh^{-1}(1)$ for $g=0$. Since the $\beta$ term in the
energy will in general depend on $g$,
we can't know the value of $B$ at finite $g$ {\it a priori}, so in practice it
must be left as a free counterterm
parameter to be determined by requiring that
the final answer be consistent with Lorentz invariance.

More generally, we can regard $B$ as a free
parameter of the lattice model, which we don't necessarily
have to insist is Lorentz covariant. Taking $B$ large enough
lifts the $P^-$ of any multi-open string intermediate state above all the
$g=0$ closed string energy
eigenstates that survive the continuum limit\footnote{
In effect with $B$ this large one imposes confinement on the
free theory! The crucial issue is then whether or not confinement
survives as $B$ is reduced to a value that restores Lorentz invariance.}. Then
the severe infrared divergences in loop diagrams caused by the open string
tachyons are removed, making the multi-loop expansion
well defined, albeit with a loss of Lorentz covariance.
For numerical studies we should therefore
calculate for general $B$ large enough to remove instabilities;
and only at the end of the calculation would we scan for a
value of $B$ which restores Lorentz covariance, if possible.

But there is no {\it a priori} guarantee that these two
counterterms
can remove all Lorentz violating
artifacts from this lattice worldsheet construction.
For instance, the authors of \cite{bardakcit}
developed a lightcone worldsheet formalism which mapped
the planar diagrams of gauge theories in lightcone gauge
to a worldsheet system with exactly the features of the open
string planar diagrams just described. String coordinates were
employed, but their dynamics were topological in the sense
that all but a single zero mode decoupled in the path history
sum. In this way the worldsheet diagrams were designed to
yield precisely the ``bare'' Feynman diagrams of the gauge theory.
Digitization of this worldsheet in the manner of GT amounted to
a scheme for cutting off the UV and IR divergences of these
Feynman diagrams. To test the reliability of this cutoff,
the authors of \cite{chakrabarti1} calculated the one loop
diagrams contributing to the scattering of glue by glue
regulated by the GT worldsheet lattice. They found that
the artificial divergences associated with lightcone quantization
could indeed be absorbed in the bulk or boundary worldsheet
counterterms. Further the divergences associated with charge
renormalization (asymptotic freedom) had the correct
coefficients. Unfortunately, this was not the end of the story.
There remained gauge violating terms that could only be cancelled by:
(1) a divergent gluon self mass, (2) a finite wave function
renormalization, (3) a finite adjustment to the three gluon
function, and (4) a finite constant adjustment to the 4 gluon
function. Indeed these are precisely the adjustments generally
required when a physical cutoff is employed in loop calculations
\cite{johnsonqed}.
All of these adjustments, being polynomials
in the external momenta, are consistent with locality. The bottom
line is that the lightcone worldsheet lattice as
a regulator of gauge theory diagrams is no better
than other physical cutoffs. Unfortunately this means that
we should expect the necessary counterterms to proliferate
with the inclusion of multiloop diagrams.

In this paper we begin to explore whether the GT lattice does a reliable
(or at least better) job regulating open string planar diagrams than
it does with field theory planar diagrams.
Does keeping $\alpha^\prime>0$ control the proliferation of
counterterms? Is it possible that the bulk and boundary counterterms
will be sufficient by themselves?
We start with the simplest self energy diagram: the one loop correction to
the closed string propagators. This process involves only a single
intermediate open string state and hence has the singularity structure
of a tree diagram. Since the corrections to
the open string propagator involve
complications associated with the multi-string intermediate sates
and worldsheet boundaries, we choose to
defer the open string analysis to a subsequent paper,
and we restrict our attention here to the corrections to the
closed string propagator.

Because of the tree structure of the closed string self-energy diagrams,
the sum over
$K$, the number of time steps that the intermediate open string exists,
converges for both the closed string tachyon and graviton, even with $B=B_0$.
Nonetheless the calculations of Sections
\ref{sec_closedstring_numerics} and  \ref{sec_dbranes_numerics}
(with $B=B_0$)
will establish Lorentz violations in both the tachyon and graviton.
Fortunately, these Lorentz violations disappear if the calculations are done
holding $B>B_0$, taking $B\to B_0$ only at the end of the calculation.
In any case, this is the only way to make sense of multi-loop diagrams,
so we don't think its necessity at one loop is a drawback.

The article is organized as follows.
In Section \ref{sec_closedstring} we obtain
explicit formulas for the one loop energy shifts of the
closed string ground state (tachyon) and the closed
string graviton state.
Each term in the $K$ summand involves determinants of $M\times M$
overlap matrices. These formulas are evaluated and
analyzed with the help of \texttt{Mathematica}
in Section \ref{sec_closedstring_numerics}\footnote{For the interested reader, we provide the evaluation code and a sample of our analysis in a \texttt{Mathematica} file accompanying the source format of this article on the arXiv.}.
In Section \ref{sec_dbranes} we extend the formulas
to include open strings ending on D$p$-branes,
and similarly perform their numerical evaluation in
Section \ref{sec_dbranes_numerics}. In this
case, instead of an energy shift, the one loop correction gives the
amplitude for a closed string scattering off the D$p$-brane.
Additional discussion is given in the concluding Section \ref{conclusion}.
Several appendices collect
a number of technical results, including normal mode
expansions, determinant formulas, and overlap matrices, that
are useful for the discussion in the main text. We also include additional evidence for the robustness of our numerical results.

\section{Closed String Self-Energy}\label{sec_closedstring}
Before calculating the self-energy diagram on the lattice,
we briefly recall the known
expression for the continuum self-energy. In the language of conformal
field theory we need the amplitude with
two closed string vertex operators, say at 1 and $\infty$,
on the complex plane from which a disk of radius $q<1$
has been excised. Consulting
for example \cite{gsw}, we find the result for the closed string
tachyon self-energy:
\bea
-\Delta P^-&=&{C\over2P^+}\int_0^1{dq\over q^3}(1-q^2)^2
\label{secft}
\eea
which is obviously seriously divergent at $q=0$: there are quadratic
and logarithmic divergences in the $q$ integration. To get further insight
into the fate of these divergences, we do the path integral in
lightcone parameterization, using the conformal
transformation methods of \cite{kacdrum,mandelstamdet}. The (still seriously
divergent) result in $D$ spacetime dimensions is
\bea
-\Delta P^-&=&
C^\prime P^+\int_0^\infty dT
\left[{2\pi \over P^+\sinh(\pi TT_0/P^+)}\right]^{(D-2)/8}
\label{selc}
\eea
where $T$ is the length of the slit on the lightcone Mandelstam
diagram, which is the total $ix^+$ over which the intermediate
open string propagates. The factor of $P^+$ is just the
result of integrating the $\sigma$ independent integrand
over vertical location of the slit
$0<\sigma<P^+$. When $D=26$ this expression reduces to
(\ref{secft}) which can be seen with the change of integration variable
\bea
q={1-e^{-\pi T T_0/P^+}\over1+e^{-\pi T T_0/P^+}}.
\eea
We can get a rough idea of what we should get from the lattice calculation
by simply discretizing $T=aK$ and $P^+=aMT_0$. Then the discretization
of (\ref{selc}) reads:
\bea
-a\Delta P^-&=&
\left(C^\prime T_0^{-2}(aT_0)^{(26-D)/8}\right)M\sum_{K=1}^\infty
\left[{2\pi \over M\sinh(\pi K/M)}\right]^{(D-2)/8}
\label{selcdisc}
\eea
Of course discretizing the result of a continuum calculation is
not the same as doing the discretized calculation from the beginning,
but it at least can serve to guide the eye. For example one
feature we immediately see from (\ref{selcdisc}) is that
the quadratic divergence seen in (\ref{secft}) is expected
to be reflected in the lattice calculation as a linear term in
$M$ in $P^-$, which can be absorbed in the bulk counterterm
described in the introduction. We turn next to the
actual lattice calculation of the self-energy.

\begin{figure}[ht]
\begin{center}
\includegraphics[width=4in]{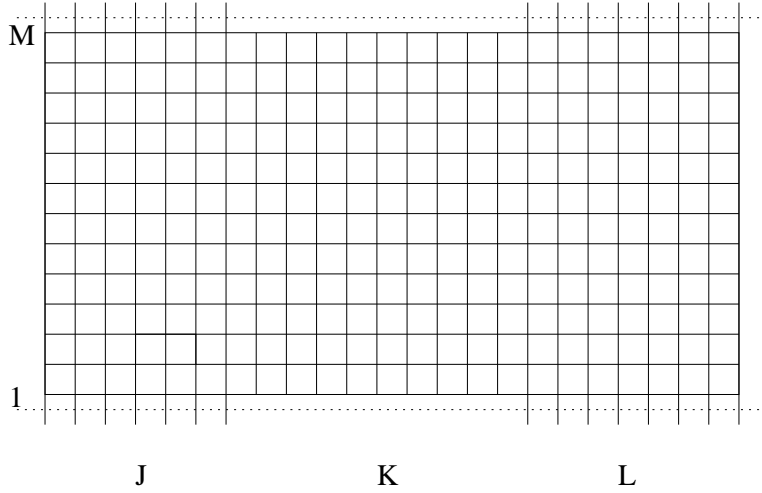}
\caption{Lattice worldsheet for the closed string self-energy. The dotted
lines are identified. There are $K-1$ missing links.}
\label{closedselattice}
\end{center}
\end{figure}
We start with the expression for the summand of the one loop correction
to the closed string propagator, depicted in Fig.~\ref{closedselattice}.
It is a product of factors, one for each of the $D-2$ transverse coordinates
${\bfs x}_i^j$. In the following we will display only 1 of the factors,
calling its coordinate $x_i^j$:
\bea
\VEV{N+1,\{x^f\}|0,\{x^i\}}^{closed}(K,J)&=&\int dx^K_idx^L_i
\VEV{L,\{x^f\}|0,\{x^L\}}^{closed}\VEV{K,\{x^L\}|0,\{x^K\}}^{open}\nonumber\\
&&
\VEV{J,\{x^K\}|0,\{x^i\}}^{closed}
e^{-{T_0}[(x^L_M-x^L_1)^2+(x^K_M-x^K_1)^2]/4}\\
&&\hskip-2in={\cal D}^{closed}(J){\cal D}^{open}(K)
{\cal D}^{closed}(L)\int dx^K_idx^L_i
e^{iW-(K-1)B_0-{T_0}[(x^L_M-x^L_1)^2+(x^K_M-x^K_1)^2]/4}\nonumber\eea
where $J+K+L=N+1$, and $B_0=(1/2)\sinh^{-1}1$ is the counter term
which removes the boundary contribution to the free open
string $P^-$. Regarding the intermediate open string as
a closed string with a row of missing links, we see that
$K-1$ is the number of missing links, which we have taken to
be the ones linking site 1 to site $M$. The open and closed string free
propagators are defined in Appendix~\ref{propagators}. The factors
${\cal D}$, which are the corresponding propagators for vanishing
values for the initial and final coordinates, are related to
determinants of the worldsheet discretized Laplacian, and are
defined in Appendix~\ref{determinants}.

Energy eigenvalues are determined by identifying exponential time
($\tau=ix^+$)
dependence in the closed string propagator $e^{-P^-\tau}\to
e^{-a(N+1)P^-}$. In perturbation theory $P^-=P^-_0+\Delta P^-$
and
\bea
e^{-a(N+1)P^-}&=&e^{-a(N+1)P_0^-}\left[1-a(N+1)\Delta P^-+\cdots\right]
\eea
so we identify $-a\Delta P^-$ as the coefficient of $N+1$ in the
one loop correction to the closed string propagator. The factor $N+1$
is associated with one of the sums over the creation and destruction
times of the intermediate open string. In practice the factor is removed
by using time translation invariance to fix, say, the creation
time, and summing only over the destruction time.

If we desire to use this formalism to calculate the corrections to
excited energy levels, it is necessary to deal with the fact
that then the summand over $K$ grows exponentially with $K$.
A convenient approach is to add a term $-K\epsilon$ to the exponent,
which can be thought of as adding $\epsilon$ to $B_0$. Since all
of the physically meaningful excited states have an excitation
energy of order $1/M$,
as long as $\epsilon>0$ is independent of $M$ this is enough to
tame the exponential divergence. Actually, since $B_0$ is
a necessary counterterm which is expected to receive corrections
in perturbation theory, we cannot know {\it a priori} what its value
should be. All we know is $B=(1/2)\sinh^{-1} 1+\mathcal{O}(g^2)$. Thus in
the context of nonperturbative studies of the worldsheet path
integral, it should be taken as a parameter to be tuned at the
end of the calculation to ensure Lorentz covariance.

For some states, such as the ground state and graviton state, the absence
of lower energy open string states that couple means that the $K$ sum
converges without the need for $\epsilon$. However, as we shall see
later, because the convergence is then only like
$e^{-K/M}$, ultraviolet divergences become entangled with infrared divergences,
which then leads to violations of Lorentz invariance since $M\propto P^+$.
Introducing an $\epsilon>0$ prevents these artifacts from entering.
Thus, we should keep $\epsilon>0$ even in those fortuitous cases where it
isn't strictly necessary for convergence.
\subsection{Correction to the Closed String Ground Energy}
Since there is only one ground state, it is uniquely singled out
by taking $J,L\to\infty$, in which case it suffices to simplify matters
by setting $x^i=x^f=0$.
Then $iW$ is simplified to
\bea
iW&=&-{T_0\over2}\bigg[q_0^{K2}\left({1\over J}+{1\over K}\right)
+q_0^{L2}\left({1\over K}+{1\over L}\right)-2q_0^Lq_0^K{1\over K}
\nonumber\\
&&+\sum_m(q_{cm}^{K2}+q_{sm}^{K2})\sinh\lambda_m^c\coth J\lambda_m^c
+\sum_m(q_{cm}^{L2}+q_{sm}^{L2})\sinh\lambda_m^c\coth L\lambda_m^c
\nonumber\\
&&+\sum_m(q_{om}^{L2}+q_{om}^{K2})\sinh\lambda_m^o\coth K\lambda_m^o
-2\sum_m q_{om}^{L}q_{om}^{K}{\sinh\lambda_m^o\over
\sinh K\lambda_m^o}\bigg]
\eea
In this equation $\lambda^o_m=2\sinh^{-1}\sin(m\pi/2M)$ and
$\lambda^c_m=2\sinh^{-1}\sin(m\pi/M)$ are the discrete time versions
of the normal mode frequencies for the open and closed strings,respectively.
Correspondingly, the $q$'s are the normal mode
coordinates of the open or closed string, defined in
Appendix~\ref{normalmodes}.
Note that when it makes no essential difference,
we shall restrict $M$ to be odd to keep the description of the
closed string modes as simple as possible.

We change integration variables to the closed string
normal mode coordinates $q_0,q_{cm},q_{sm}$ for both $K$ and $L$.
The Jacobian for this variable change is unity, and we can express the $q_{om}$
in terms of the closed string modes as follows:
\bea
q_{om}&=&\cases{q_{cm/2} &$m$ {\rm even}\cr&\cr
{2\over M}\sum_{m^\prime=1}^{(M-1)/2}q_{sm^\prime}U_{mm^\prime} &
$m$ {\rm odd}\cr}
\eea
where the overlap matrix $U_{mm^\prime}$ is defined in Appendix~\ref{overlap}.
We also need
\bea
x_M-x_1&=&-2\sqrt{2\over M}\sum_{m^\prime=1}^{(M-1)/2}q_{sm^\prime}
\sin{m^\prime\pi\over M}
\eea
Then
\bea
&&\hskip-24ptiW-{T_0\over4}\left[(x^L_M-x^L_1)^2+(x^K_M-x^K_1)^2\right]=\nonumber\\
&&-{T_0\over2}\Bigg[q_0^{K2}\left({1\over J}+{1\over K}\right)
+q_0^{L2}\left({1\over K}+{1\over L}\right)-2q_0^Lq_0^K{1\over K}
\nonumber\\
&&+\sum_m q_{cm}^{K2} \sinh\lambda_m^c(\coth J\lambda_m^c+\coth K\lambda_m^c)
+\sum_m q_{cm}^{L2}\sinh\lambda_m^c(\coth L\lambda_m^c+\coth K\lambda_m^c)
\nonumber\\
&&-2\sum_m q_{cm}^{L}q_{cm}^{K}{\sinh\lambda_m^c\over
\sinh K\lambda_m^c}
+\sum_m q_{sm}^{K2}\sinh\lambda_m^c\coth J\lambda_m^c
+\sum_m q_{sm}^{L2}\sinh\lambda_m^c\coth L\lambda_m^c
\nonumber\\
&&+\sum_{m~{\rm odd}}(q_{om}^{L2}+q_{om}^{K2})
\sinh\lambda_m^o\coth K\lambda_m^o
-2\sum_{m~{\rm odd}} q_{om}^{L}q_{om}^{K}{\sinh\lambda_m^o\over
\sinh K\lambda_m^o}\nonumber\\
&&+{4\over M}\sum_{m^\prime,m^{\prime\prime}}
(q^K_{sm^\prime}q^K_{sm^{\prime\prime}}
+q^L_{sm^\prime}q^L_{sm^{\prime\prime}})\sin{m^\prime\pi\over M}
\sin{m^{\prime\prime}\pi\over M}\Bigg]
\label{action}
\eea
Because of the equality $q_{o2m}=q_{cm}$, the integration over
$q_0$ and the $q_{cm}$ precisely implements closure on these modes.
This means that the result of those integrations is just
the contribution of those modes to ${\cal D}^{closed}(N+1)$.
Consulting Appendix~\ref{determinants} for the various ${\cal D}$'s,
the outcome of the integration over $q_0$ and the
$q_{cm}$ can be written
\bea
&&\hskip-24pt\VEV{N+1,\{x^f\}|0,\{x^i\}}^{closed}\nonumber\\
&&={\cal D}^{\rm closed}_{\cos}(N+1)
{\cal D}_{\sin}^{closed}(J){\cal D}_{\rm odd}^{open}(K)
{\cal D}_{\sin}^{closed}(L)\int dq_{sm}^Kdq_{sm}^L
e^{iW^\prime-(K-1)B_0}\nonumber\\
&&={\cal D}^{\rm closed}(N+1)
{{\cal D}_{\sin}^{closed}(J){\cal D}_{\rm odd}^{open}(K)
{\cal D}_{\sin}^{closed}(L)\over{\cal D}_{\sin}^{closed}(N+1)}
\int dq_{sm}^Kdq_{sm}^L
e^{iW^\prime-(K-1)B_0}\\
&&\hskip-24pt
iW^\prime=-{T_0\over2}\Bigg[\sum_m(q_{sm}^{K2})
\sinh\lambda_m^c\coth J\lambda_m^c
+\sum_m(q_{sm}^{L2})\sinh\lambda_m^c\coth L\lambda_m^c
\nonumber\\
&&+\sum_{m^\prime,m^{\prime\prime}}
(q^K_{sm^\prime}q^K_{sm^{\prime\prime}}
+q^L_{sm^\prime}q^L_{sm^{\prime\prime}})
\bigg({4\over M^2}\sum_{m~{\rm odd}}U_{mm^\prime}U_{mm^{\prime\prime}}
\sinh\lambda_m^o\coth K\lambda_m^o\nonumber\\
&&\hskip.5in+{4\over M}\sin{m^\prime\pi\over M}
\sin{m^{\prime\prime}\pi\over M}\bigg)-{8\over M^2}\sum_{m^\prime,m^{\prime\prime}}q^K_{sm^\prime}q^L_{sm^{\prime\prime}}\sum_{m~{\rm odd}}
U_{mm^\prime}U_{mm^{\prime\prime}}{\sinh\lambda_m^o\over
\sinh K\lambda_m^o}\Bigg]
\eea
and we remind the reader that there are $D-2$ such factors.
To isolate the shift in a specific energy level, we need to identify
the exponential behavior in $J,L$ as they approach infinity.

For the ground state energy shift, it is sufficient
to directly take $J,L\to\infty$ with $K$ fixed.
Then the $\coth J,\coth L\to1$ and
\bea
{\cal D}^{\rm closed}_{\sin}(N+1)&\to&e^{-(N+1)\sum_{m=1}^{(M-1)/2}
\lambda^c_m/2}\left({T_0\over2\pi}\right)^{(M-1)/4}
\prod_{m=1}^{(M-1)/2}\sqrt{2\sinh\lambda^c_m}\\
{{\cal D}^{\rm closed}_{\sin}(J)
{\cal D}^{\rm closed}_{\sin}(L)\over{\cal D}^{\rm closed}_{\sin}(N+1)}
&\to&e^{K\sum_{m=1}^{(M-1)/2}
\lambda^c_m/2}\left({T_0\over2\pi}\right)^{(M-1)/4}
\prod_{m=1}^{(M-1)/2}\sqrt{2\sinh\lambda^c_m}\eea
Change integration variables to
${\bar q}_m=q_{sm}\sqrt{(T_0/\pi)\sinh\lambda^c_m}$.
And we have
\bea
&&\hskip-.4in\VEV{N+1,\{x^f\}|0,\{x^i\}}^{closed}\nonumber\\
&&\to{\cal D}^{\rm closed}(N+1)
{{\cal D}_{\rm odd}^{open}(K)e^{K\sum_{m=1}^{(M-1)/2}
\lambda^c_m/2}
\over\prod_m\sqrt{(T_0/\pi)\sinh\lambda^c_m}}
\int d{\bar q}_{m}^Kd{\bar q}_{m}^L
e^{iW^{\prime\prime}-(K-1)B_0}\nonumber\\
&&\to{\cal D}^{\rm closed}(N+1)
{e^{K\sum_{m=1}^{M-1}
(\lambda^c_m-\lambda^o_m)/2}
\over\prod_{m=1,{\rm odd}}^{M-1}\sqrt{1-e^{-2K\lambda^o_m}}}\prod_{m=1}^{M-1}
\sqrt{\sinh\lambda^o_m\over\sinh\lambda^c_m}
\int d{\bar q}_{m}^Kd{\bar q}_{m}^L
e^{iW^{\prime\prime}-(K-1)B_0}
\eea
where $iW^{\prime\prime}$ is defined below. It can be shown that
\bea
\prod_{m=1}^{M-1}\sinh\lambda^o_m&=& 2^{-(M-1)}
\sqrt{M{\sinh2M\sinh^{-1}1\over\sinh2\sinh^{-1}1}}\nonumber\\
&=& 2^{-(M-1)}
\sqrt{M{\sinh M\sinh^{-1}1\cosh M\sinh^{-1}1\over\sqrt{2}}}\\
\prod_{m=1}^{M-1}\sinh\lambda^c_m&=& 2^{-(M-1)}
M\sinh M\sinh^{-1}1
\eea
so, restoring all $D-2$ factors in the amplitude and
summing only over $K$, the time interval spanned by the open
string propagator, we infer
\bea
&&\hskip-64pt -a\Delta P^-_{G,closed}=M\sum_{K}\left[{\VEV{N+1,\0\}|0,\{0\}}^{closed}
\over{\cal D}^{\rm closed}(N+1)}\right]^{D-2} \nonumber\\
&=&
M\sum_{K=1}^\infty\left[{e^{K\sum_{m=1}^{M-1}
(\lambda^c_m-\lambda^o_m)/2-(K-1)B_0}
\over\prod_{m=1,odd}^{M-1}\sqrt{1-e^{-2K\lambda^o_m}}}
\left({\coth M\sinh^{-1}1\over M\sqrt{2}}\right)^{1/4}
\int d{\bar q}_{m}^Kd{\bar q}_{m}^L
e^{iW^{\prime\prime}}\right]^{D-2}\\
iW^{\prime\prime}&=&-{\pi\over2}\bigg[\sum_m({\bar q}_{m}^{K2}
+{\bar q}_{m}^{L2})
+\sum_{m^\prime,m^{\prime\prime}}
{{\bar q}^K_{m^\prime}{\bar q}^K_{m^{\prime\prime}}
+{\bar q}^L_{m^\prime}{\bar q}^L_{m^{\prime\prime}}\over
\sqrt{\sinh\lambda^c_{m^\prime}\sinh\lambda^c_{m^{\prime\prime}}}}\nonumber\\
&&\quad\bigg({4\over M^2}\sum_{m~{\rm odd}}U_{mm^\prime}U_{mm^{\prime\prime}}
\sinh\lambda_m^o\coth K\lambda_m^o+{4\over M}\sin{m^\prime\pi\over M}
\sin{m^{\prime\prime}\pi\over M}\bigg)\nonumber\\
&&\quad-{8\over M^2}
\sum_{m^\prime,m^{\prime\prime}}
{{\bar q}^K_{sm^\prime}{\bar q}^L_{sm^{\prime\prime}}\over
\sqrt{\sinh\lambda^c_{m^\prime}
\sinh\lambda^c_{m^{\prime\prime}}}}\sum_{m~{\rm odd}}
U_{mm^\prime}U_{mm^{\prime\prime}}{\sinh\lambda_m^o\over
\sinh K\lambda_m^o}\bigg]\nonumber\\
&\equiv&-\pi\left[\sum_{m^\prime,m^{\prime\prime}}
({\bar q}^K_{m^\prime}{\bar q}^K_{m^{\prime\prime}}
+{\bar q}^L_{m^\prime}{\bar q}^L_{m^{\prime\prime}})A_{m^\prime
m^{\prime\prime}}+2\sum_{m^\prime,m^{\prime\prime}}
{\bar q}^K_{m^\prime}{\bar q}^L_{m^{\prime\prime}}B_{m^\prime
m^{\prime\prime}}\right]
\eea
With the definitions on the last line
\bea
\int d{\bar q}_{m}^Kd{\bar q}_{m}^L
e^{iW^{\prime\prime}}&=&{\det}^{-1/2}\pmatrix{A&B\cr B&A\cr}
={\det}^{-1/2}(A+B)\ {\det}^{-1/2}(A-B).\eea
The last equality follows because the eigenvectors of
$\pmatrix{A&B\cr B&A\cr}$ can be taken to be of the form
$\pmatrix{v_\pm\cr\pm v_\pm\cr}$ where $v_\pm$ is an
eigenvector of $A\pm B$.
Finally we summarize the result
\bea
-a\Delta P^-_{G,closed}&=&\nonumber\\
&&\hskip-.8in
M\sum_{K=1}^\infty\left[\left({\coth M\sinh^{-1}1\over M\sqrt{2}}\right)^{1/4}
{e^{K\sum_{m=1}^{M-1}
(\lambda^c_m-\lambda^o_m)/2-(K-1)B_0}
\over\prod_{m=1,odd}^{M-1}\sqrt{1-e^{-2K\lambda^o_m}}}
{\det}^{-1/2}\pmatrix{A&B\cr B&A\cr}\right]^{D-2}\label{shift_closed_groundstate}\\
A_{m^\prime m^{\prime\prime}}&=&
{\delta_{m^\prime m^{\prime\prime}}\over2}+{2\over M}{\sin(m^\prime\pi/M)
\sin(m^{\prime\prime}\pi/M)\over\sqrt{\sinh\lambda^c_{m^\prime}
\sinh\lambda^c_{m^{\prime\prime}}}}\nonumber\label{A_matrix}\\
&&\hskip1.8in+{2\over M^2}\sum_{m~{\rm odd}}
{U_{mm^\prime}U_{mm^{\prime\prime}}\sinh\lambda^o_m\coth K\lambda^o_m
\over\sqrt{\sinh\lambda^c_{m^\prime}
\sinh\lambda^c_{m^{\prime\prime}}}}\\%
B_{m^\prime m^{\prime\prime}}&=&-{2\over M^2}\sum_{m~{\rm odd}}
{U_{mm^\prime}U_{mm^{\prime\prime}}\sinh\lambda^o_m
\over\sqrt{\sinh\lambda^c_{m^\prime}
\sinh\lambda^c_{m^{\prime\prime}}}\sinh K\lambda^o_m}\label{B_matrix}\\
(A\pm B)_{m^\prime m^{\prime\prime}}&=&
{\delta_{m^\prime m^{\prime\prime}}\over2}+{2\over M}{\sin(m^\prime\pi/M)
\sin(m^{\prime\prime}\pi/M)\over\sqrt{\sinh\lambda^c_{m^\prime}
\sinh\lambda^c_{m^{\prime\prime}}}}\nonumber\\
&&\hskip1in+{2\over M^2}\sum_{m~{\rm odd}}
{U_{mm^\prime}U_{mm^{\prime\prime}}\sinh\lambda^o_m
\over\sqrt{\sinh\lambda^c_{m^\prime}
\sinh\lambda^c_{m^{\prime\prime}}}}\left[\tanh{K\lambda^o_m\over2}
\right]^{\pm1}
\eea
\subsection{Correction to the Graviton Energy}
For the shift in excited energy levels, we need to identify the
non-leading exponential behaviors in the free closed string propagator.
Since we also want to pick out a specific spin, it is iimportant to
work with the complete amplitude, including all $D-2$ factors:
\bea
\VEV{N+1,\{q^f\}|0,\{q^i\}}_{\rm total}^{closed}&=&
\left[{\cal D}^{closed}(N+1)\right]^{D-2}
e^{iW_{\rm total}^{closed}}\\
&&\hskip-2in iW_{\rm total}^{closed}=-{T_0\over2}
\Bigg[{({\bfs q}_{0,N+1}-{\bfs q}_{0,0})^2\over N+1}+
\sum_{m=1}^{M-1}\sinh\lambda^c_m\Bigg(({\bfs q}_{m,N+1}^2
+{\bfs q}_{m,0}^2)\coth(N+1)\lambda^c_m\nonumber\\
&&\qquad -{2{\bfs q}_{m,N+1}\cdot{\bfs q}_{m,0}\over\sinh(N+1)\lambda^c_m}
\Bigg)\Bigg]
\eea
where ${\bfs q}$ denotes a $D-2$ dimensional vector.
Here we identify the $m<M/2$ modes with cosine modes and the
$m>M/2$ modes with sine modes.
The graviton state on the lattice has energy $2\lambda^c_1$ above the
ground state energy, and involves only the 1 modes.
Also since it is a symmetric traceless $O(D-2)$ tensor,
its contribution to the closed string propagator resides in the
second order term in the expansion of
\bea
\exp\Bigg[{T_0({\bfs q}^c_{1,N+1}\cdot{\bfs q}^c_{1,0}
+{\bfs q}^s_{1,N+1}\cdot{\bfs q}^s_{1,0})
\sinh\lambda^c_1\over\sinh(N+1)\lambda^c_1}
\Bigg]&\sim&\nonumber\\
&&\hskip-3in1+{2T_0({\bfs q}^c_{1,N+1}\cdot{\bfs q}^c_{1,0}
+{\bfs q}^s_{1,N+1}\cdot{\bfs q}^s_{1,0})\sinh\lambda^c_1}
e^{-(N+1)\lambda^c_1}\\
&&\hskip-2in+{2T_0^2({\bfs q}^c_{1,N+1}\cdot{\bfs q}^c_{1,0}
+{\bfs q}^s_{1,N+1}\cdot{\bfs q}^s_{1,0})^2
\sinh^2\lambda^c_1}e^{-2(N+1)\lambda^c_1}\nonumber
\eea
as $N+1\to\infty$.
The first term (the 1) propagates an $O(D-2)$ scalar, the second term
propagates an $O(D-2)$ vector, and the third term propagates a combination
of a traceless symmetric tensor, and antisymmetric tensor and a scalar.
Since all closed string states must be cyclically invariant, the second
(vector) term is projected out of the spectrum. But, in any case,
for the shift in the graviton energy, we may simply drop the first
two terms, and keep only the symmetric traceless, cyclically invariant
part of the third term.

To identify the contribution of the cyclically symmetric states to the
third term we consider the new coordinates
\bea
{\bfs q}_1^{\pm}&\equiv&{\bfs q}_1^{c}\pm i{\bfs q}_1^{s}
\eea
which acquire the factor $e^{\pm2\pi i/M}$ under a cyclic transformation
of one step. To make a cyclically invariant combination, we must have
equal numbers of $+$ and $-$ factors:
\bea
{\bfs q}^c_{1,N+1}\cdot{\bfs q}^c_{1,0}
+{\bfs q}^s_{1,N+1}\cdot{\bfs q}^s_{1,0}&=&{1\over2}
({\bfs q}^+_{1,N+1}\cdot{\bfs q}^-_{1,0}
+{\bfs q}^-_{1,N+1}\cdot{\bfs q}^+_{1,0})\\
({\bfs q}^c_{1,N+1}\cdot{\bfs q}^c_{1,0}
+{\bfs q}^s_{1,N+1}\cdot{\bfs q}^s_{1,0})^2&=&{1\over4}
[({\bfs q}^+_{1,N+1}\cdot{\bfs q}^-_{1,0})^2
+({\bfs q}^-_{1,N+1}\cdot{\bfs q}^+_{1,0})^2]\nonumber\\
&&+{1\over2}
{\bfs q}^+_{1,N+1}\cdot{\bfs q}^-_{1,0}\
{\bfs q}^-_{1,N+1}\cdot{\bfs q}^+_{1,0}\nonumber\\
&\to&{1\over2}
{\bfs q}^+_{1,N+1}\cdot{\bfs q}^-_{1,0}\
{\bfs q}^-_{1,N+1}\cdot{\bfs q}^+_{1,0}
={1\over2}
{q}^{+k}_{1,N+1}{q}^{-l}_{1,N+1}\
{q}^{-k}_{1,0}{q}^{+l}_{1,0}\nonumber
\eea
where the last line shows the only contribution that survives the
cyclic symmetry requirements. To see the $SO(D-2)$ content write
\bea
{q}^{-k}_{1,0}{q}^{+l}_{1,0}&=&q_{1,0}^{ck}q_{1,0}^{cl}
+q_{1,0}^{sk}q_{1,0}^{sl}
+i(q_{1,0}^{ck}q_{1,0}^{sl}
-q_{1,0}^{sk}q_{1,0}^{sk})
\eea
The third (imaginary) term gives the contribution of the
anti-symmetric tensor, whereas the first two (real) terms
give a symmetric tensor, which can further be decomposed into
a traceless symmetric tensor and a scalar:
\bea
q_{1,0}^{ck}q_{1,0}^{cl}
+q_{1,0}^{sk}q_{1,0}^{sl}=\left[q_{1,0}^{ck}q_{1,0}^{cl}
+q_{1,0}^{sk}q_{1,0}^{sl}
-{\delta_{kl}\over D-2}({\bfs q}_{1,0}^{c2}+{\bfs q}_{1,0}^{s2})\right]
+{\delta_{kl}\over D-2}({\bfs q}_{1,0}^{c2}+{\bfs q}_{1,0}^{s2})
\eea
The quantity in square brackets on the right represents the contribution of
the graviton in all spin configurations. To identify the graviton energy
shift it is sufficient to simply pick one polarization with $k\neq l$ so
that the trace subtraction drops out.
\bea
\left[\exp\Bigg\{{T_0({\bfs q}^c_{1,N+1}\cdot{\bfs q}^c_{1,0}
+{\bfs q}^s_{1,N+1}\cdot{\bfs q}^s_{1,0})
\sinh\lambda^c_1\over\sinh(N+1)\lambda^c_1}
\Bigg\}\right]_{cyc~inv}&\sim&\nonumber\\
&&\hskip-2in1+{T_0^2
{q}^{+k}_{1,N+1}{q}^{-l}_{1,N+1}\
{q}^{-k}_{1,0}{q}^{+l}_{1,0}\sinh^2\lambda^c_1}e^{-2(N+1)\lambda^c_1}
+\cdots\\
&&\hskip-4.5in=1+{T_0^2
(q_{1,N+1}^{ck}q_{1,N+1}^{cl}
+q_{1,N+1}^{sk}q_{1,N+1}^{sl})(q_{1,0}^{ck}q_{1,0}^{cl}
+q_{1,0}^{sk}q_{1,0}^{sl})\sinh^2\lambda^c_1}e^{-2(N+1)\lambda^c_1}\\
&&\hskip-4.2in+T_0^2(q_{1,N+1}^{ck}q_{1,N+1}^{sl}
-q_{1,N+1}^{sk}q_{1,N+1}^{cl})(q_{1,0}^{ck}q_{1,0}^{sl}
-q_{1,0}^{sk}q_{1,0}^{cl})\sinh^2\lambda^c_1e^{-2(N+1)\lambda^c_1}
+\cdots
\eea
The first 1 term can be dropped in the calculation of the
graviton and also the anti-symmetric tensor energy shifts, since it
contributes only for scalar states (the tachyonic ground state and
the massless dilaton. Then we can take $J,L\to\infty$ in
\bea
&&\hskip-18pt
\VEV{N+1,\{x^f\}|0,\{x^i\}}_{\rm total}^{Graviton}(K,J)\nonumber\\
&&\sim T_0^4\sinh^4\lambda^c_1
\left[{\cal D}^{closed}(J){\cal D}^{open}(K){\cal D}^{closed}(L)\right]^{D-2}
e^{-2(L+J)\lambda^c_1}\nonumber\\
&&
\int dq^K_m dq^L_i[q_{f,1}^{ck}q_{f,1}^{cl}
+q_{f,1}^{sk}q_{f,1}^{sl}][q_{L,1}^{ck}q_{L,1}^{cl}
+q_{L,1}^{sk}
q_{L,1}^{sl}][q_{K,1}^{ck^\prime}q_{K,1}^{cl^\prime}
+q_{K,1}^{sk^\prime}q_{K,1}^{sl^\prime}][q_{i,1}^{ck^\prime}q_{i,1}^{cl^\prime}
+q_{i,1}^{sk^\prime}q_{i,1}^{sl^\prime}]\nonumber\\
&&\hskip2.5in
e^{iW_{\rm total}-(K-1)(D-2)B_0-{T_0}[({\bfs x}^L_M-{\bfs x}^L_1)^2
+({\bfs x}^K_M-{\bfs x}^K_1)^2]/4}
\eea
The sum over $K$ of this expression should be compared to the
free closed string propagator for the graviton
\bea
\left[{\cal D}(N+1)\right]^{D-2}T_0^2\sinh^2\lambda^c_1e^{-2(N+1)\lambda^c_1}
[q_{f,1}^{ck}q_{f,1}^{cl}+q_{f,1}^{sk}q_{f,1}^{sl}]
[q_{i,1}^{ck}q_{i,1}^{cl}
+q_{i,1}^{sk}q_{i,1}^{sl}]
\eea
to read off the graviton energy shift:
\bea
&&MT_0^2\sinh^2\lambda^c_1\sum_{K=1}^\infty
\left[{{\cal D}^{closed}(J){\cal D}^{open}(K){\cal D}^{closed}(L)
\over{\cal D}(N+1)}\right]^{D-2}e^{2K\lambda^c_1-(K-1)(D-2)B_0}
\nonumber\\
&&\int dq^K_m dq^L_i[q_{L,1}^{ck}q_{L,1}^{cl}
+q_{L,1}^{sk}
q_{L,1}^{sl}][q_{K,1}^{ck^\prime}q_{K,1}^{cl^\prime}
+q_{K,1}^{sk^\prime}q_{K,1}^{sl^\prime}]
e^{iW_{\rm total}-{T_0}[({\bfs x}^L_M-{\bfs x}^L_1)^2+({\bfs x}^K_M
-{\bfs x}^K_1)^2]/4}\nonumber\\
&&=-{a\over2}(\delta_{kk^\prime}\delta_{ll^\prime}+
\delta_{kl^\prime}\delta_{lk^\prime})\Delta P^-_{Graviton}+C\delta_{kl}
\delta_{k^\prime l^\prime}
\eea
where the $C$ term contributes to the dilaton energy shift. (If $C=0$
the graviton and dilaton remain degenerate.) Here $iW$ is the same expression
(\ref{action}) that we used in the evaluation of the
ground state energy shift.
A simple way to isolate the graviton shift is to simply choose
index values for which the $C$ term decouples. For example,
take $k=k^\prime=1$ and $l=l^\prime=2$:
\bea\label{shift_graviton}
-a\Delta P^-_{Graviton}&=&2MT_0^2\sinh^2\lambda^c_1\sum_{K=1}^\infty
\left[{{\cal D}^{closed}(J){\cal D}^{open}(K){\cal D}^{closed}(L)
\over{\cal D}(N+1)}\right]^{D-2}e^{2K\lambda^c_1-(K-1)(D-2)B_0}
\nonumber\\
&&\hskip-.4in\int dq^K_m dq^L_m[q_{L,1}^{c1}q_{L,1}^{c2}
+q_{L,1}^{s1}
q_{L,1}^{s2}][q_{K,1}^{c1}q_{K,1}^{c2}
+q_{K,1}^{s1}q_{K,1}^{s2}]
 e^{iW_{\rm total}-{T_0}[({\bfs x}^L_M-{\bfs x}^L_1)^2
+({\bfs x}^K_M-{\bfs x}^K_1)^2]/4}
\nonumber\\
&=&2MT_0^2\sinh^2\lambda^c_1\sum_{K=1}^\infty
\left[{{\cal D}^{closed}(J){\cal D}^{open}(K){\cal D}^{closed}(L)
\over{\cal D}(N+1)}\right]^{D-2}e^{2K\lambda^c_1-(K-1)(D-2)B_0}
\nonumber\\
&&\hskip-.6in\int dq^K_m dq^L_m[q_{L,1}^{c1}q_{K,1}^{c1}q_{L,1}^{c2}
q_{K,1}^{c2}+q_{L,1}^{s1}q_{K,1}^{s1}q_{L,1}^{s2}q_{K,1}^{s2}]
 e^{iW_{\rm total}-{T_0}[({\bfs x}^L_M-{\bfs x}^L_1)^2
+({\bfs x}^K_M-{\bfs x}^K_1)^2]/4}
\nonumber\\
&\equiv&2MT_0^2\sinh^2\lambda^c_1\sum_{K=1}^\infty
\left[{{\cal D}^{closed}(J){\cal D}^{open}(K){\cal D}^{closed}(L)
\over{\cal D}(N+1)}\right]^{D-2}e^{2K\lambda^c_1-(K-1)B_0}
\nonumber\\
&&\hskip-.6in(\VEV{q_{L,1}^{c1}q_{K,1}^{c1}}^2
+\VEV{q_{L,1}^{s1}q_{K,1}^{s1}}^2)
\int dq^K_m dq^L_m
 e^{iW_{\rm total}-{T_0}[({\bfs x}^L_M-{\bfs x}^L_1)^2
+({\bfs x}^K_M-{\bfs x}^K_1)^2]/4}
\eea
where in the second and third forms
we take advantage of the fact that the integration
is over independent Gaussians, so the language of correlations
reflected in the $\VEV{\cdots}$ notation is appropriate.
The correlator of cosine modes is just that of the free closed string
and is easily shown to be
\bea\label{propagator1}
\VEV{q_{L,1}^{c1}q_{K,1}^{c1}}&=&{\sinh L\lambda^c_1\ \sinh J\lambda^c_1
\over T_0\sinh\lambda^c_1\ \sinh(K+J+L)\lambda^c_1}\to
{1\over2T_0\sinh\lambda^c_1}e^{-K\lambda^c_1}
\eea
in the limit $J,L\to\infty$.

The correlator of sine modes is of course more complicated because they
involve the nontrivial overlap of the closed and open string modes.
\bea
\VEV{q_{L,1}^{s1}q_{K,1}^{s1}}&=&{\pi\over T_0\sinh\lambda^c_1}
\VEV{{\bar q}_{L,1}^{s1}{\bar q}_{K,1}^{s1}}\\
\VEV{{\bar q}_{L,1}^{s1}{\bar q}_{K,1}^{s1}}&=&{\int
d{\bar q}_{K,m}d{\bar q}_{L,m}
{\bar q}_{L,1}^{s}{\bar q}_{K,1}^{s}e^{iW^{\prime\prime}}\over
\int
d{\bar q}_{K,m}d{\bar q}_{L,m}e^{iW^{\prime\prime}}}
\eea
Recall that
$iW^{\prime\prime}=-\pi {\bar q}^T\pmatrix{A&B\cr B&A\cr}{\bar q}$
so adding a source term
$J^T{\bar q}$, we complete the square to evaluate
\bea
\VEV{e^{J^T{\bar q}}}&=&\exp\left\{{1\over4\pi}
J^T\pmatrix{A&B\cr B&A\cr}^{-1}J\right\}
\eea
With the definition
\bea
\pmatrix{A&B\cr B&A\cr}^{-1}&=&\pmatrix{A^\prime&B^\prime\cr
B^\prime&A^\prime\cr}\\
A^\prime&=&(A-BA^{-1}B)^{-1}={1\over2}\left((A+B)^{-1}+(A-B)^{-1}\right)\nonumber\\
B^\prime&=&(B-AB^{-1}A)^{-1}={1\over2}\left((A+B)^{-1}-(A-B)^{-1}\right),
\eea
we then have
\bea\label{propagator2}
\VEV{q_{L,1}^{s1}q_{K,1}^{s1}}&=&{1\over 2T_0\sinh\lambda^c_1}B^\prime_{1,1}
\eea

\section{Numerical Analysis}\label{sec_closedstring_numerics}
\subsection{Closed String Ground State}

Here we will perform a numerical study of the 1-loop shift in the
ground state energy for the closed string in $D=26$ dimensions,
with the help of \texttt{Mathematica}. In particular,
we will rescale the energy shift (\ref{shift_closed_groundstate}) by an overall factor of $M$ (notice also $\Delta P^-_{G,closed}$ and $\delta P^-_K$ differ by a minus sign),
\bea\label{rescaled_DeltaP}
-\frac{a \Delta P^-_{G,closed}}{M}&\equiv&\sum_{K=1}^\infty \delta P^-_K=\nonumber\\
&&\hskip-.8in
=\sum_{K=1}^\infty\left[\left({\coth (M\sinh^{-1}1)\over M\sqrt{2}}\right)^{1/4}
{e^{K\sum_{m=1}^{M-1}
(\lambda^c_m-\lambda^o_m)/2-(K-1)B_0}
\over\prod_{m=1,odd}^{M-1}\sqrt{1-e^{-2K\lambda^o_m}}}
{\det}^{-1/2}\pmatrix{A&B\cr B&A\cr}\right]^{24}\,,
\eea
where $A$ and $B$ are given by (\ref{A_matrix}),(\ref{B_matrix}), as it will turn out that this yields a finite quantity for $M\to \infty$, which is just
what is expected from the bulk term in $\Delta P^-$.
\begin{figure}
\begin{center}
\includegraphics[width=0.9\textwidth]{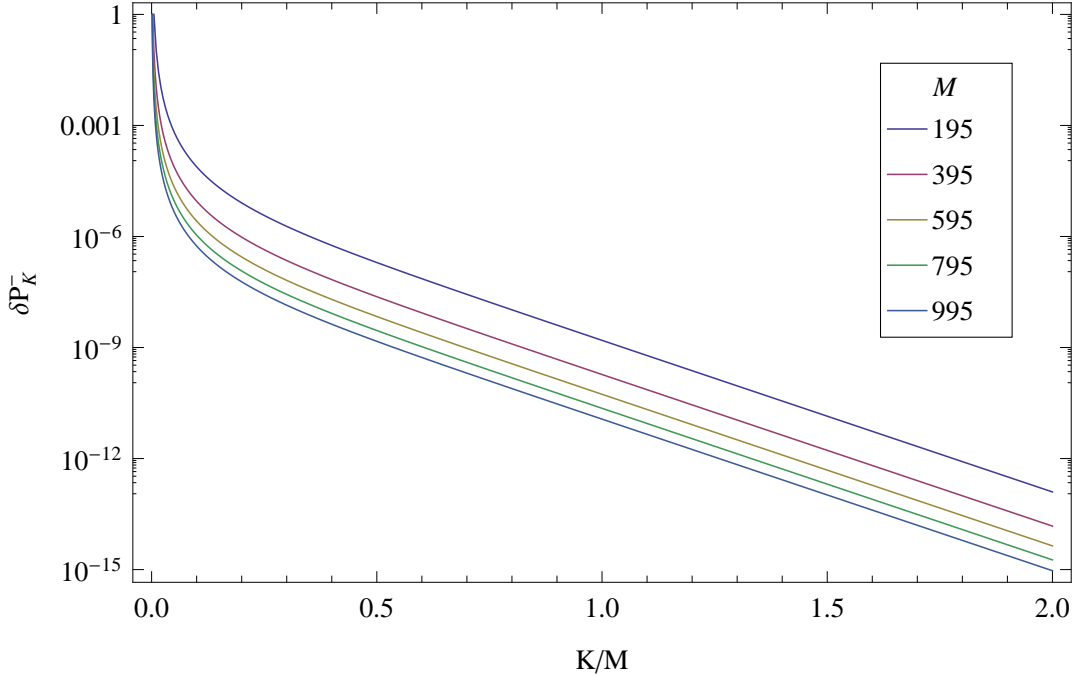}
\caption{\small Log-linear plot for the summand of the rescaled ground state energy shift (\ref{rescaled_DeltaP}), with each curve exhibiting its dependence on $K$ for a fixed value of $M$. We have rescaled the horizontal axis to $K/M$ in order to demonstrate that for a large enough value of the latter, $\log \delta P^-_K$ develops the same slope. The plot also justifies our choice $K_{max}=2M$ for the cutoff in the $K$ sum.}
\label{fig:DeltaP_K_Log}
\end{center}
\end{figure}
It is instructive to start by investigating the dependence of the summand $\delta P^-_K$ on $M$ and $K$, as a means to also set a reasonable cutoff $K_{max}\ge K$ in the sum. For fixed $M$, $\delta P^-_K$ decreases rapidly for increasing $K$, and for $K\gg M$ it becomes proportional to $e^{-9.428 K/M}$ times an $M$-dependent factor. This fact is evident in figure \ref{fig:DeltaP_K_Log},
which also indicates that $K_{max}=2M$ is a sufficiently large cutoff,
given that the largest term in the sum is $\delta P_1^-=1$ for any $M$, and $\delta P_{K_{max}}\le 10^{-12}$ for $M\ge195$, precisely indicating the accuracy of our cutoff.

\begin{figure}
\includegraphics[width=0.49\textwidth]{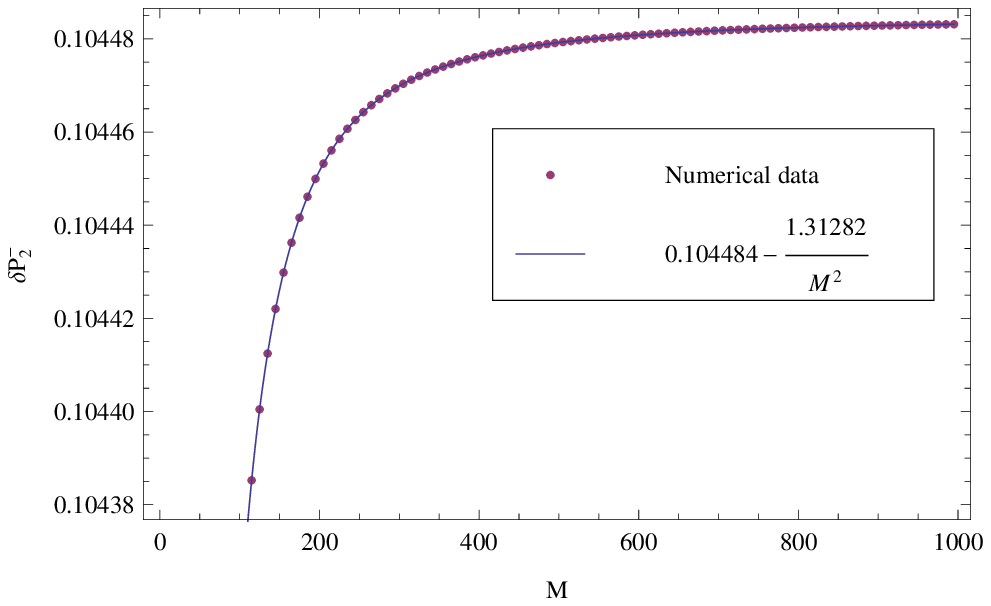}
\includegraphics[width=0.49\textwidth]{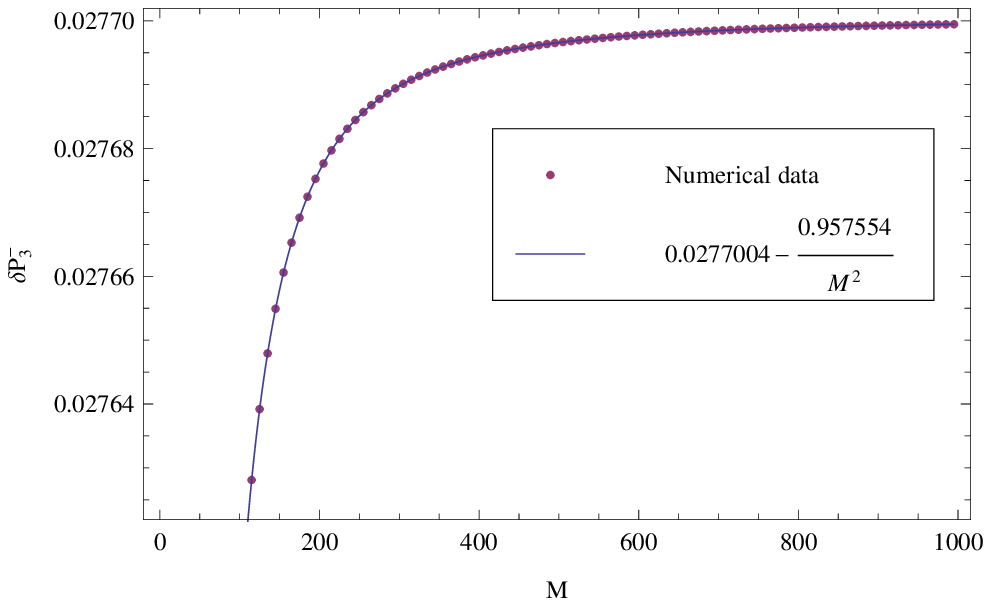}
\includegraphics[width=0.49\textwidth]{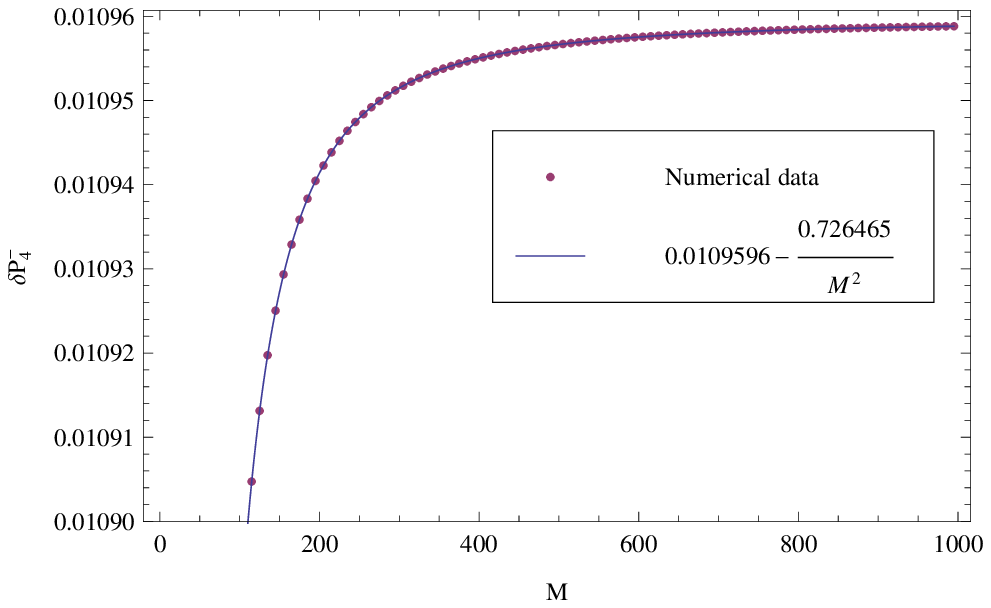}
\includegraphics[width=0.49\textwidth]{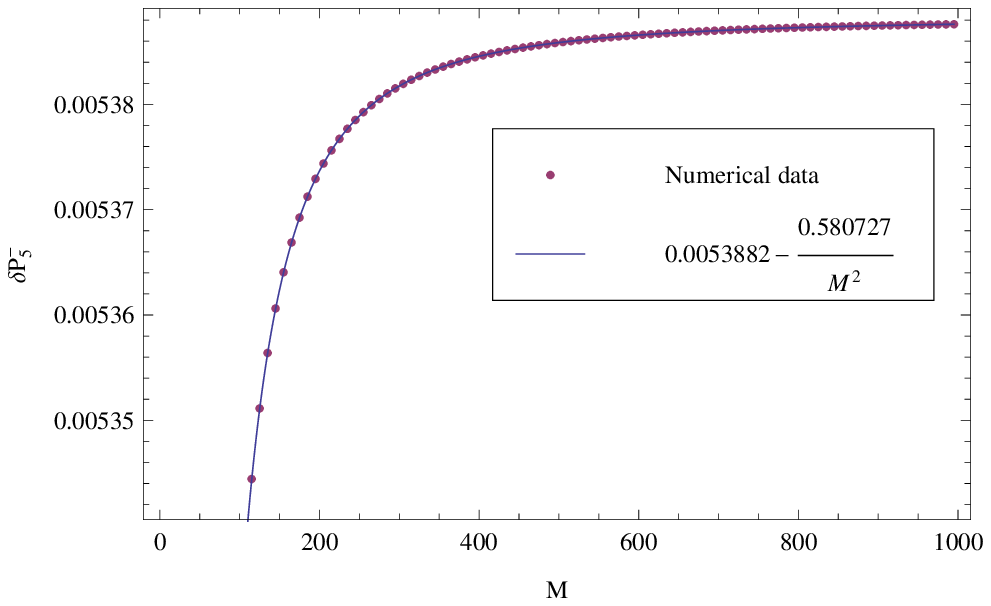}
\caption{\small Summand for the rescaled ground state energy shift as a function of $M$ for $K=2,3,4,5\,\ll M$, including fits of the form (\ref{deltaP_M_expansion}).}
\label{fig:DeltaP_K}
\end{figure}

For fixed $K\ll M$, we also find a good fit to
\beq\label{deltaP_M_expansion}
\delta P^-_K=c_1^K+\frac{c_2^K}{M^2}+\mathcal{O}(\frac{1}{M^3})
\eeq
where roughly $c_1^K\sim K^{-3}$ and $c_2^K\sim K^{-1}$. The particular examples $K=2,3,4,5$ are presented in figure \ref{fig:DeltaP_K}. Since this behavior changes as $K$ becomes comparable to $M$, it suggests that the sum over $K$ should give rise to a $\log M/M^2$ term due to
\beq
\sum_{K=1}^M c_2^K\sim H(M)\sim \log M \textrm{ for } M\gg1\,,\label{harmonicsum}
\eeq
where $H(M)$ is the $M$-th harmonic number. Similarly the sum of $c_1^K$ yields harmonic numbers of order 3, whose large $M$ expansion suggests the absence of an $1/M$ term.

We should also note that in both regimes we examined, the general structure of the dependence on $M$ and $K$ is correctly captured by the discretized version of the continuum amplitude (\ref{selcdisc}). Denoting the summand of the latter (up to a proportionality factor) with a prime in order to avoid confusion, we have for $D=26$
\bea
\delta P^{-\prime}_K=\left(\frac{\pi}{M \sinh(\pi K/M)}\right)^3&=&\cases{\left(\frac{2\pi}{M}\right)^3 e^{-3\pi K/M}+\mathcal{O}(e^{-5\pi K/M}) &$K\gg M$\cr&\cr
\frac{1}{K^3}-\frac{\pi^2}{2K M^2}+\mathcal{O}(\frac{1}{M^4}) &
$K\ll M$\cr}
\eea
For $K\gg M$ there is also approximate agreement in the value of the exponent, although for the $K\ll M$ expansions more detailed comparison of the coefficients of $\delta P^{-\prime}_K$ and $\delta P^{-}_K$ reveals that they are not simply proportional to each other.

Armed with this intuition, we proceed to the numerical calculation
of (\ref{rescaled_DeltaP}), summed up to $K_{max}=2M$,
and for values of $M$ ranging from 5 to 995 in steps of 10. We fit the
generated data for different subintervals between $M\in [195,995]$
to ensure that $M$ is sufficiently large and to test
the stability of our fits, and also calculate
the value of $R^2$ as an estimate of their goodness.
We find that indeed the fit
\beq\label{self_energy_fit}
-\frac{a \Delta P^-_{G,closed}}{M}=c_1+c_2 \frac{1}{M^2}+c_3\frac{\log M}{M^2}
\eeq
with
\beq\label{self_energy_fit_coefficients}
c_1=1.158863267\pm 3\cdot 10^{-9}\,,\quad c_2=2.799\pm 0.011\,,\quad c_3=-2.800\pm0.002
\eeq
matches excellently with the data, with the values of the coefficients varying only mildly when fitting different subintervals in $M$ (the error estimates are precisely taking this interval dependence into account).
\begin{figure}
\begin{center}
\includegraphics[width=0.83\textwidth]{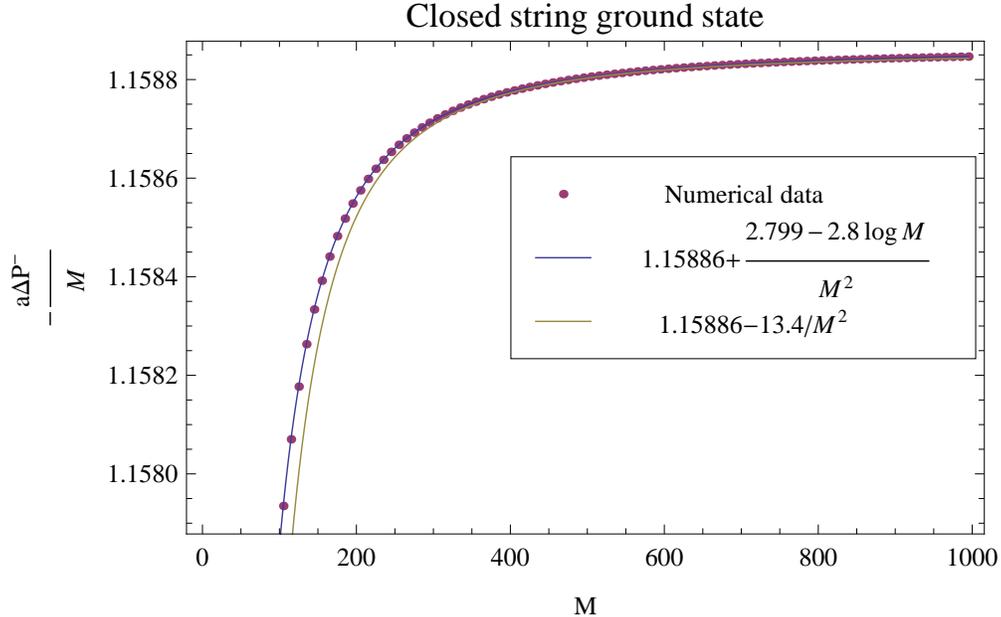}
\caption{\small Rescaled ground state energy shift as a function of $M$. We present fits with and without a $\log M$ term, in order to demonstrate the necessity of the latter for agreement with the data.}
\label{fig:DeltaP}
\end{center}
\end{figure}

Our main finding of this section, (\ref{self_energy_fit}),
is plotted against the  numerical data in figure \ref{fig:DeltaP}.
We've also included the fit with $c_3=0$ to show its insufficiency
in accurately describing the data. As far as the fit
with $c_2=0$ is concerned, it leads to values of $c_3$
which may differ up to $7\%$ depending on the interval of the fit,
and generally one should also expect a constant multiplying $M$
inside the logarithm. As additional evidence that the $M$-dependence
is indeed correctly captured by (\ref{self_energy_fit}),
we also mention that when fitting the entire interval
$M\in [195,995]$, the value of $R^2$ differs from 1 by a mere
$3\cdot 10^{-11}$, whereas for the $c_2=0$ and $c_3=0$ cases
the differences are $6\cdot 10^{-5}$ and $0.002$ respectively.
Finally, our expectations for the absence of an $1/M$ term is
confirmed by the fact that its inclusion yields unnaturally
small values for its coefficient, and does not substantially improve the fit.

As described in the introduction, the bulk counterterm
can be chosen to cancel the contribution to $P^-$ proportional to $M$,
and what is left gives the physically significant contribution.
Lorentz invariance requires that this residuum behave at large $M$ as $1/M$,
since $\Delta m^2=2M aT_0 (\Delta P^--{\rm Bulk~Term})$.
Our results (\ref{self_energy_fit}),(\ref{self_energy_fit_coefficients})
contradict this requirement because of the $\ln M$ dependence. Taken literally,
the result implies a logarithmically divergent self mass:
$\ln M=\ln P^+/aT_0=\ln(1/a)+\ln(P^+/T_0)$, which is to be expected from
the $dq/q$ behavior in the covariant expression for the self-energy.
As is well known this divergence can be absorbed in a renormalization of
the Regge slope parameter $\alpha^\prime=1/2\pi T_0$. But the $\ln P^+$
signifies a noncovariant finite part.

The origin of this $\ln M$ factor can be traced to the sum over $K$
of the $1/K$ dependence we saw in the summand when $K<<M$. The lightcone
lattice has cutoff the logarithmic UV divergence (small $K$), but the
large $K$ behavior is cut off at $K=\mathcal{O}(M)$, because the level spacing
is of order $1/M$, so $\sum_K(1/K)=\mathcal{O}(\ln M)$.
As mentioned in the introduction, the presence of the
$B_0$ counterterm offers a way to interpret the lattice calculation
that avoids the difficulty. By adding a small positive constant
$B_0\to B_0+\epsilon$ the cutoff on the $K$ sum becomes $1/\epsilon$
instead of $M$, and the residuum will behave as $(1/M)\ln(1/\epsilon)$
which is still divergent as $\epsilon\to0$, but remains compatible
with Lorentz invariance. Then the $\ln(1/\epsilon)$ can be absorbed in
a redefinition of $T_0$ before taking $M\to\infty$. As we shall see in
the next subsection, this same interpretation leads to a zero self-energy for the graviton state.

\subsection{Graviton}

\begin{figure}
\includegraphics[width=0.48\textwidth]{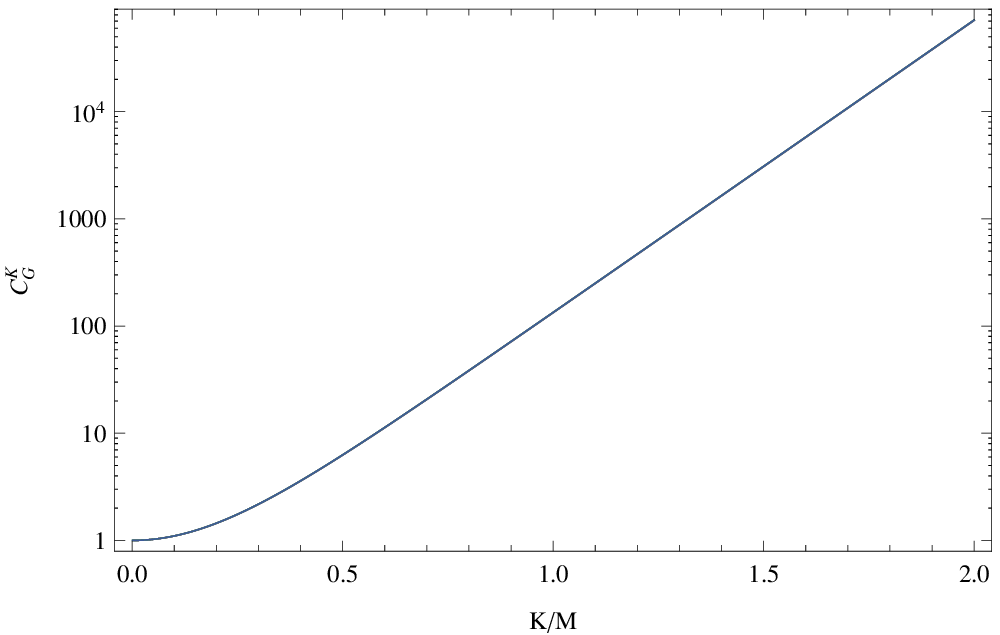}
\includegraphics[width=0.48\textwidth]{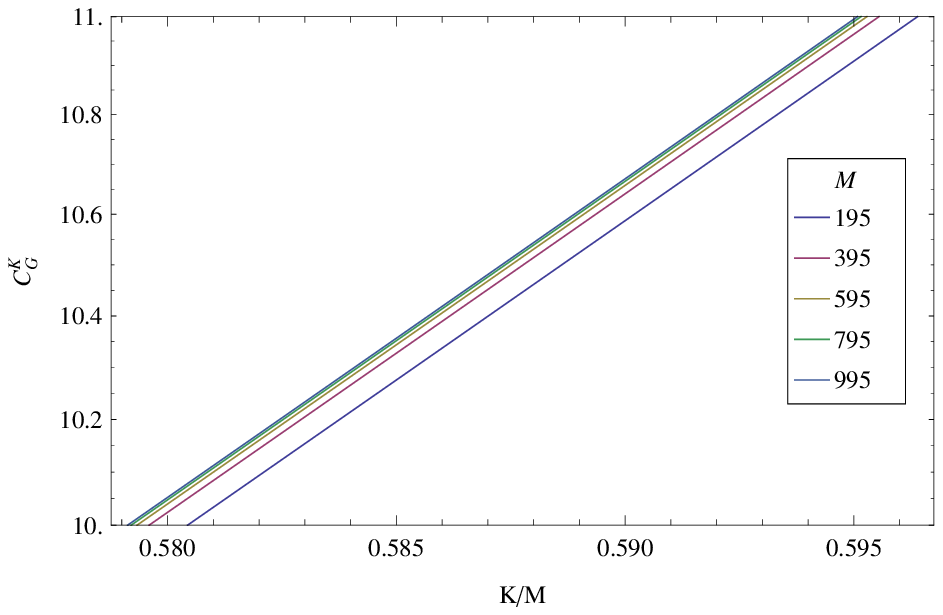}
\caption{\small Coefficient $C_G^K$ in the rescaled graviton energy shift (\ref{shift_graviton_rescaled}) as a function of $K/M$, with each curve corresponding to varying $K$ for a fixed value of $M$. On the left hand side we plot the entire range of summation for $K/M$, it is evident that curves for different $M$ values are indistinguishable. On the right hand side we zoom into a small region in order to see the mild dependence of the $\log C_G^K$ offset on $M$.}
\label{fig:C_G}
\end{figure}
\begin{figure}
\includegraphics[width=0.48\textwidth]{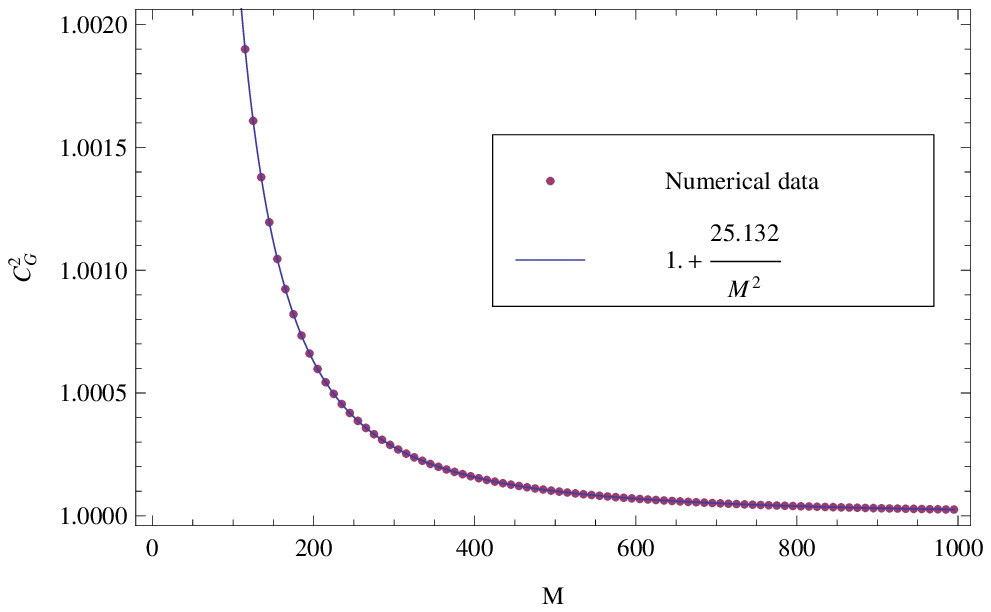}
\includegraphics[width=0.48\textwidth]{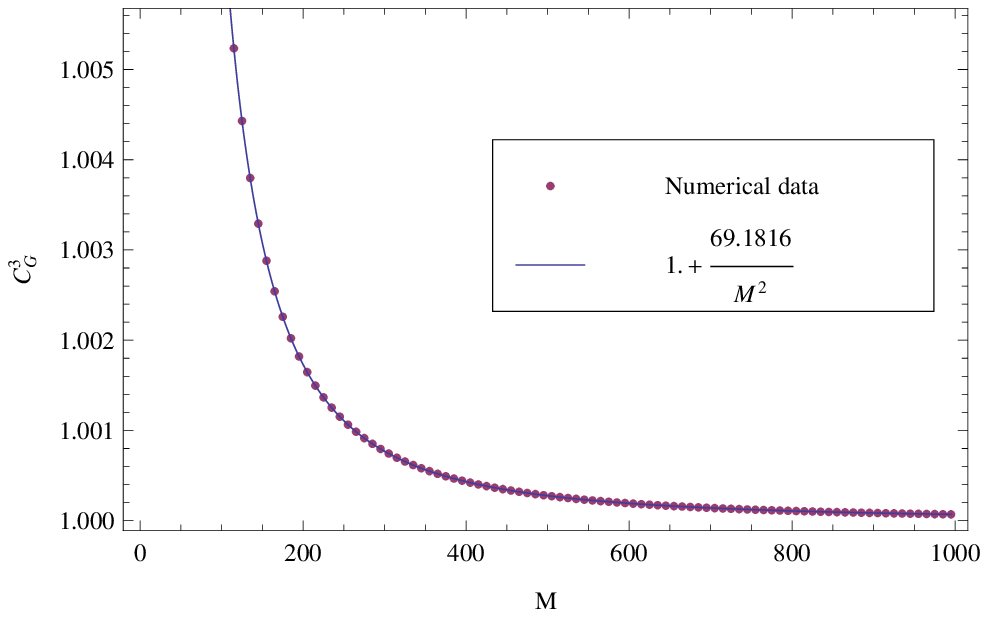}
\caption{\small Coefficient $C_G^K$ in the rescaled graviton energy shift (\ref{shift_graviton_rescaled}) as a function $M$ for fixed $K=2,3\ll M$.}
\label{fig:C_G_M}
\end{figure}
\begin{figure}
\includegraphics[width=0.48\textwidth]{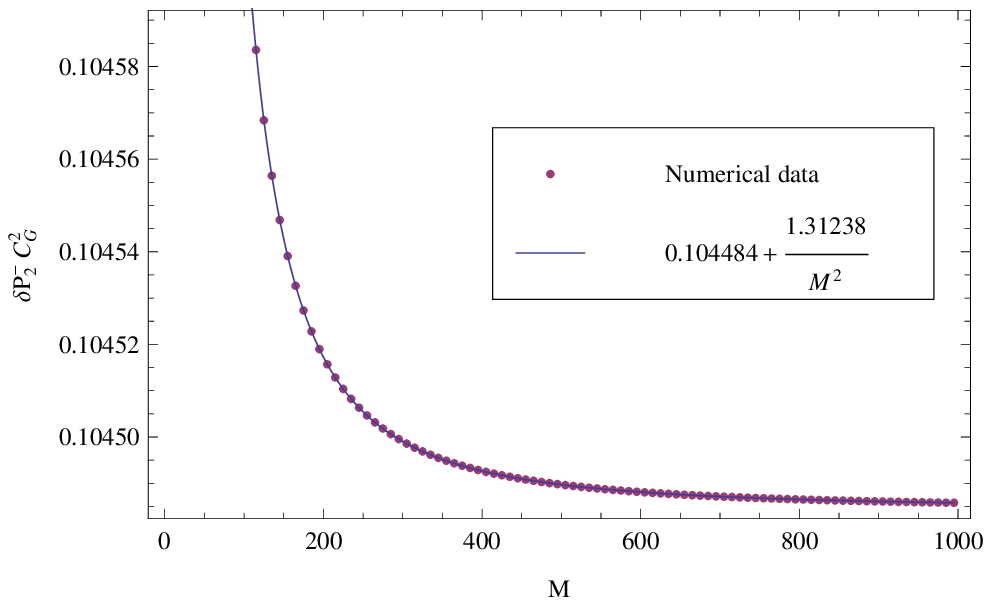}
\includegraphics[width=0.48\textwidth]{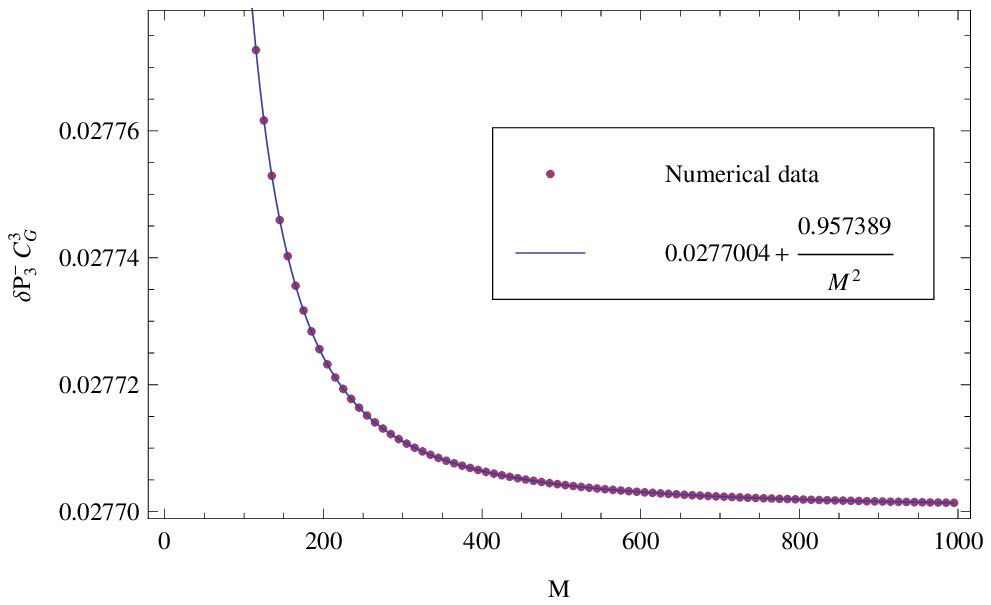}
\caption{\small $C_G^K \delta P^-_K$ as a function $M$ for fixed $K=2,3\ll M$. Comparing with figure \ref{fig:DeltaP_K}, we notice that the constant terms are equal, and the $1/M^2$ coefficients roughly opposite.}
\label{fig:C_G_DeltaP}
\end{figure}

We proceed to investigate how the lightcone lattice handles
nontachyonic states by looking at the 1-loop mass shift of
the spin-2 excitation of the closed string, representing the graviton.
Similarly to the ground state,
with the help of (\ref{propagator1}),(\ref{propagator2}),
we may rewrite (\ref{shift_graviton}) in the $J,L\to\infty$ limit as
\beq\label{shift_graviton_rescaled}
-\frac{a \Delta P^-_{Graviton}}{M}
=\frac{1}{2}\sum_{K=1}^\infty
[1+(e^{K\lambda_1^c}B^\prime_{1,1})^2]\delta P^-_K
\equiv\frac{1}{2}\sum_{K=1}^\infty (1+C_G^K)\delta P^-_K\,,
\eeq
where $\delta P^-_K$ is the summand of the rescaled ground state shift,
defined in (\ref{rescaled_DeltaP}).

As $\delta P^-_K$ has been determined in the previous section, the only additional numerical computation that has to be done is for the coefficient $C_G^K$. A preliminary analysis shows that for fixed $M$ and varying $K$ this is a rapidly increasing function which for $K\gg M$ becomes proportional to roughly $e^{6.28K/M}$, as can be seen in figure \ref{fig:C_G}. However given the behavior of $\delta P^-_K$ in the same regime, their product is guaranteed to converge, albeit more slowly. As far as the regime $K\ll M$ is concerned, we observe that $C_G^1=1$ for any $M$, and more generally $C_G^K= 1+c/M^2+\mathcal{O}(1/M^3)$. The first two nontrivial examples $K=2,3$ are plotted in figure \ref{fig:C_G_M}. The fact that the constant term is independent of $K$ and equal to one guarantees that the leading divergence for the ground state and the graviton is the same, as it should for all states.
\begin{figure}[ht]
\begin{center}
\includegraphics[width=0.83\textwidth]{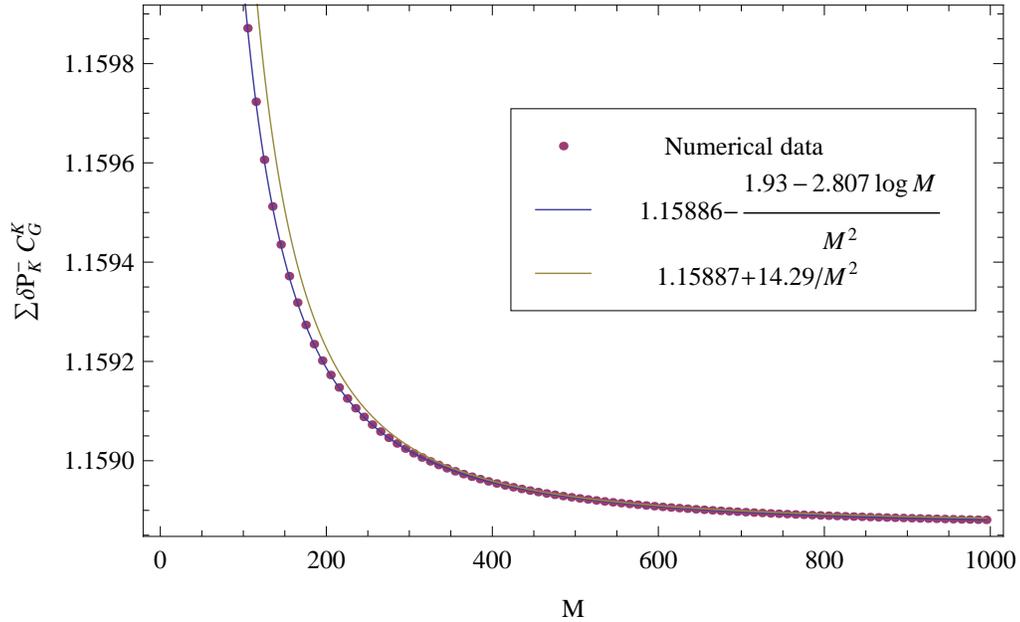}
\caption{\small $\sum_{K} C_G^K\delta P^-_K$ as a function of $M$. Again fits with and without a $\log M$ term are presented, so as to demonstrate its necessity for matching with the data.}
\label{fig:sumC_G_DeltaP}
\end{center}
\end{figure}

\begin{figure}
\begin{center}
\includegraphics[width=0.83\textwidth]{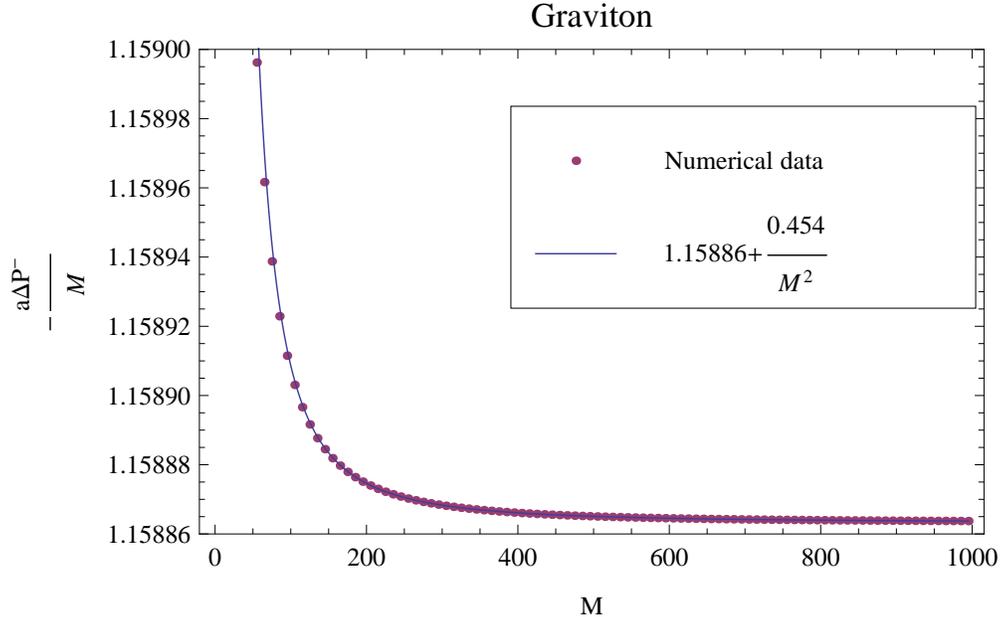}
\caption{\small Rescaled graviton energy shift as a function of $M$.}
\label{fig:DeltaPgrav}
\end{center}
\end{figure}

Before evaluating the entire sum (\ref{shift_graviton_rescaled}),
it is again useful to examine $C_G^K\delta P^-_K$
for fixed $K$. As can be seen in figure \ref{fig:C_G_DeltaP},
for individual $K\ll M$ the latter has an expansion in $M$ of the form
(\ref{deltaP_M_expansion}), where $c_1^K$ are roughly equal and $c_2^K$
are roughly opposite between the ground state and the graviton.
Then the sum in $K$ is depicted in figure
\ref{fig:sumC_G_DeltaP} similarly described by a fit of the form (\ref{self_energy_fit}), where again the coefficients $c_1$ and $c_3$ are found to be equal and opposite respectively within our margins of error, however $c_2=-1.93\pm0.04$. This is a first hint that although additional cancellations occur for the graviton, which may remove the divergent $\log M$ terms,
the lattice regularization, in the absence of the $\epsilon$
prescription, still leaves an
unphysical finite mass shift for the graviton.
\begin{figure}
\includegraphics[width=0.49\textwidth]{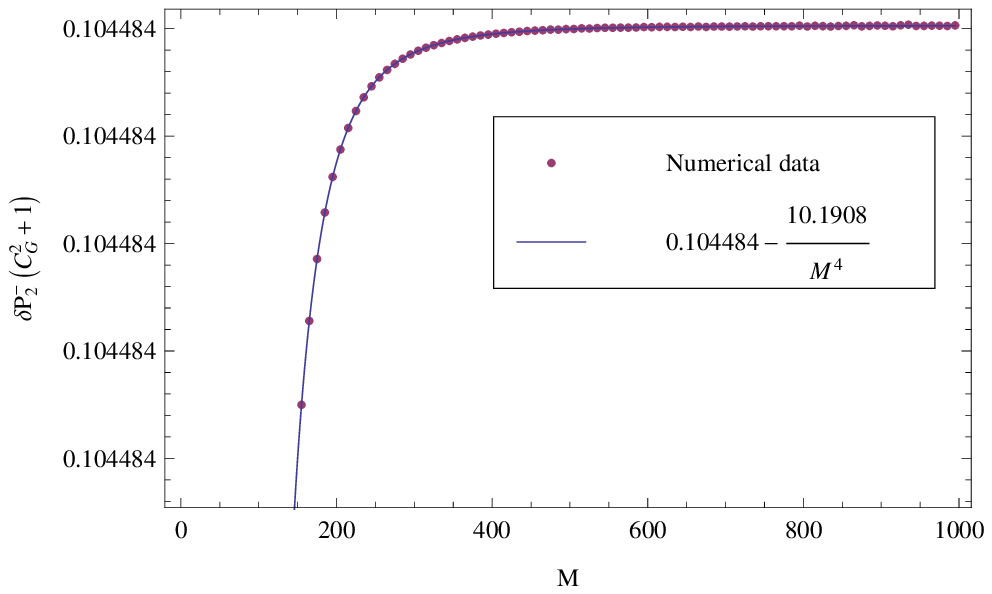}
\includegraphics[width=0.49\textwidth]{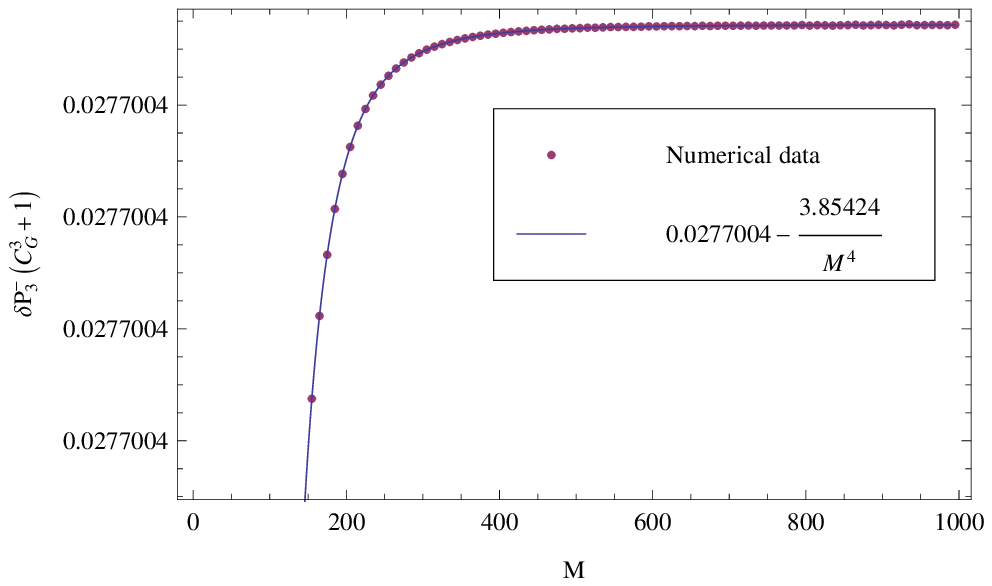}
\caption{\small Entire summand of rescaled graviton energy shift (\ref{shift_graviton_rescaled}) as a function $M$ for fixed $K=2,3\ll M$.}
\label{fig:DeltaP_summand}
\end{figure}
Moving now to the entire summand (\ref{shift_graviton_rescaled}),
from figure \ref{fig:DeltaP_summand} we infer it behaves as
\beq
\frac{1}{2}(1+C_G^K)\delta P^-_K=\tilde c_1^K+\frac{\tilde c_2^K}{M^4}
+\mathcal{O}(\frac{1}{M^5})\label{grav_summand_fit}
\eeq
with $\tilde c_1^K\simeq c_1^K\sim K^{-3}$ and $\tilde c_2^K\sim K$,
which is consistent with
\beq
\sum_{K=1}^M \tilde c_2^K\sim M^2/2 \textrm{ for } M\gg1\,,
\eeq
introducing an additional contribution that changes the
$c_2$ coefficient for the graviton, compared to the ground state.
Finally, we find that the fit for the entire
rescaled energy shift (see Fig.~\ref{fig:DeltaPgrav})
\beq
-\frac{a \Delta P^-_{Graviton}}{M}=\tilde c_1+\tilde c_2\frac{1}{M^2}\,,
\eeq
with
\beq
\tilde c_1=1.158863276\pm1.5\cdot 10^{-8}\,\quad\tilde c_2=0.454\pm0.004\,,
\eeq
is in very good agreement with the numerical data\footnote{In particular when fitting on the range $M\in[195,995]$, $R^2$ differs from 1 by $10^{-5}$. Including an additional $\log M/M^2$ term yields an unnaturally small coefficient and does not improve $R^2$ significantly.} and our previous quantitative and qualitative observations.

Our results indicate the absence
of the $\ln M$ Lorentz violating effect we found for the tachyon\footnote{
A $\ln M$ divergence in the graviton self mass could not be absorbed
in the Regge slope parameter because at zeroth
order the graviton is massless.},
but still the
$K$ sum with a cutoff of $\mathcal{O}(M)$, leads to the
undesirable conclusion that the graviton would gain a
(necessarily finite) nonzero
mass at one loop order. This is a (more subtle) violation of
Lorentz invariance.
However, with the $\epsilon$ prescription introduced to interpret
the tachyon mass shift, this difficulty is
avoided. Putting $B_0\to B_0+\epsilon$ cuts off the $K$ sum at
$\sim 1/\epsilon$, so the large $M$ expansion at fixed $\epsilon$
encounters no $1/M^2$ contribution and hence no shift in the
graviton mass. It is important to appreciate that this interpretation
requires taking the continuum limit {\it before} taking $\epsilon\to0$.

\section{D-branes}\label{sec_dbranes}
We extend the discussion to the case when several of the transverse
open string coordinates satisfy Dirichlet conditions. In current popular
terminology this is known as closed string theory in the presence of
D-branes \cite{dailp}.
To avoid confusion we will
call such coordinates ${\bfs y}_i^j$. We shall follow \cite{thorndirichlet}
in adapting the lightcone lattice to Dirichlet boundary conditions.
Starting from the closed string potential energy for one such coordinate
\bea
V_{\rm closed}&=&{T_0\over2}\sum_{i=1}^M(y_{i+1}-y_i)^2
\eea
we pass to the potential energy for a Dirichlet open string with
say $y_l=0$ by the following substitution
\bea
(y_{l+1}-y_l)^2+(y_l-y_{l-1})^2&\to&y_{l+1}^2+y_{l-1}^2+2\kappa y_l^2.
\eea
In other words we keep all $M$ degrees of freedom on the lattice.
Instead of trying to set $y_l=0$, we decouple it from the other
coordinates and give it a potential $T_0\kappa y_l^2$ that makes it
produce an energy of $\mathcal{O}(1)$ in lattice units.
This means that excitations
of $y_l$ will have
infinite energy in the continuum limit, which therefore locks
this degree of freedom in its ground state.
The normal modes for the Dirichlet open string coordinates
$q_{Dm}$ are defined in Appendix~\ref{normalmodes}. In this case
it is convenient to alter the corresponding normal mode expansion for the
closed string, which for $M$ odd are
\bea
y_k&=&{1\over\sqrt{M}}q_{0}+\sqrt{2\over M}\sum_{m=1}^{(M-1)/2}\left[q_{cm}
\cos{2m\pi k\over M}+q_{sm}\sin{2m\pi k\over M}\right].
\eea
The modification for $M$ even can be found in Appendix~\ref{normalmodes}.

We can express the open string normal modes in terms of the closed string ones
\bea
q_{{\rm D}m}&=&\cases{{\displaystyle q_{sm/2}}& for $m$ even\cr
&\cr
{\displaystyle {2\over M}\sum_{m^\prime=0}^{(M-1)/2}q_{cm^\prime}
U^{\rm D}_{mm^\prime}}& for $m$ odd\cr}
\eea
where we have defined $q_{c0}\equiv q_0/\sqrt{2}$, and $U^D_{mm^\prime}$
is given in Appendix~\ref{overlap}.

In constructing the one loop diagram, we
would like the $j=0,N+1$ sites of the open string propagator
be assigned half the closed string potential energy. We have
\bea
V^c-V^{\rm D}&=& {T_0\over2}\left((y_M-y_{M-1})^2+(y_1-y_M)^2-y_1^2-y_{M-1}^2
-2\kappa y_M^2\right)\nonumber\\
&=&-T_0y_M(y_1+y_{M-1}+(\kappa-1)y_M)\equiv -2U(y).
\eea
Thus the loop integrand is given by the product of the three propagators
times the factor $e^{U(y)}$ at each vertex.
In terms of the closed string normal modes
\bea
y_M&=&{1\over\sqrt{M}}q_{0}
+\sqrt{2\over M}\sum_{m^\prime=1}^{(M-1)/2}\hspace{-5pt}q_{cm^\prime}\equiv\sqrt{2\over M}\sum_{m^\prime=0}^{(M-1)/2}\hspace{-5pt}q_{cm^\prime}\\
y_1+y_{M-1}&=&{2\over\sqrt{M}}q_{0}
+2\sqrt{2\over M}\sum_{m^\prime=1}^{(M-1)/2}\hspace{-5pt}q_{cm^\prime}\cos{2m^\prime\pi
\over M}
\nonumber\\&\equiv&
2\sqrt{2\over M}\sum_{m^\prime=0}^{(M-1)/2}\hspace{-5pt}q_{cm^\prime}
\cos{2m^\prime\pi
\over M}\\
U(y)
&=&{T_0\over M}
\sum_{m^\prime,m^{\prime\prime}=0}^{(M-1)/2}\hspace{-5pt}q_{cm^\prime}q_{cm^{\prime\prime}}
\left[\kappa-1+\cos{2m^\prime\pi\over M}
+\cos{2m^{\prime\prime}\pi\over M}\right]
\eea
where we defined $q_{c0}\equiv q_0/\sqrt{2}$.
Then the one loop correction to the closed string propagator is
\bea
&&\hskip-18pt
\sum_K\int dy^K_idy^L_i e^{-(K-1)B_0^{\rm D}+U(y^K)+U(y^L)}
\VEV{L,\{y^f\}|0,\{y^L\}}^{closed}\VEV{K,\{y^L\}|0,\{y^K\}}^{\rm D}
\nonumber\\&&\hskip4.5in\VEV{J,\{y^K\}|0,\{y^i\}}^{closed}
\nonumber\\
&&\phantom{a}
\nonumber\\
&&\hskip1in=\sum_K{\cal D}^{closed}(J){\cal D}^{\rm D}(K)
{\cal D}^{closed}(L)\int dy^K_idy^L_i
e^{iW-(K-1)B^{\rm D}_0+U(y^K)+U(y^L)}
\nonumber
\eea
where as before we display only one factor for the Dirichlet coordinate $y$.
Note that we expect $B_0^{\rm D}\neq B_0$ because of the different
boundary conditions.
\subsection{Closed String Tachyon Scattering off D-brane}
For the case where $y_i=y_f=0$ we have, for each Dirichlet coordinate $y$, a term
\bea
iW+U(y^K)+U(y^L)
&=&-{T_0\over2}\Bigg[q_0^{K2}{1\over J}
+q_0^{L2}{1\over L}
\nonumber\\&&\hskip-1in
+\sum_m(q_{cm}^{K2}+q_{sm}^{K2})\sinh\lambda_m^c\coth J\lambda_m^c
+\sum_m(q_{cm}^{L2}+q_{sm}^{L2})\sinh\lambda_m^c\coth L\lambda_m^c
\nonumber\\&&\hskip-1in
+\sum_m(q_{{\rm D}m}^{L2}+q_{Dm}^{K2})\sinh\lambda_m^{\rm D}
\coth K\lambda_m^{\rm D}
-2\sum_m q_{{\rm D}m}^{L}q_{{\rm D}m}^{K}{\sinh\lambda_m^{\rm D}\over
\sinh K\lambda_m^{\rm D}}
\nonumber\\
&&\hskip-1in-{2\over M}
\sum_{m^\prime,m^{\prime\prime}=0}^{(M-1)/2}\left(
q^K_{cm^\prime}q^K_{cm^{\prime\prime}}
+q^L_{cm^\prime}q^L_{cm^{\prime\prime}}\right)
\left(\kappa-1+\cos{2m^\prime\pi\over M}
+\cos{2m^{\prime\prime}\pi\over M}\right)
\Bigg]\nonumber
\eea
The next step is to change integration variables to the closed string
normal modes, $q_0=q_{c0}\sqrt{2},q_{cm},q_{sm}$. The Jacobian
for the change of variables $y_k\to q_0,q_{cm},q_{sm}$ is unity,
and further changing $q_0\to q_{c0}$ gives a factor $\sqrt{2}$.
 The equality
$q_{D2m}=q_{sm}$ means that integrating over the closed string
sine modes simply implements closure on these modes. Thus
we can write
\bea
&&\hskip-.7in\VEV{N+1,\{x^f\}|0,\{x^i\}}^{closed}_{1{\rm loop}}\nonumber\\
&&={\cal D}^{\rm closed}_{\sin}(N+1)
{\cal D}_{\cos}^{closed}(J){\cal D}_{\rm odd}^{\rm D}(K)
{\cal D}_{\cos}^{closed}(L)\int 2dq_{cm}^Kdq_{cm}^L
e^{iW^{{\rm D}\prime}-(K-1)B_0^{\rm D}}\nonumber\\
&&={\cal D}^{\rm closed}(N+1)
{{\cal D}_{\cos}^{closed}(J){\cal D}_{\rm odd}^{\rm D}(K)
{\cal D}_{\cos}^{closed}(L)\over{\cal D}_{\cos}^{closed}(N+1)}
\int 2dq_{cm}^Kdq_{cm}^L
e^{iW^{{\rm D}\prime}-(K-1)B_0^{\rm D}}
\eea
where
\bea
iW^{{\rm D}\prime}&=&-{T_0\over2}\Bigg[q_0^{K2}{1\over J}
+q_0^{L2}{1\over L}
+\sum_m q_{cm}^{K2}\sinh\lambda_m^c\coth J\lambda_m^c
+\sum_m q_{cm}^{L2}\sinh\lambda_m^c\coth L\lambda_m^c
\nonumber\\&&
+\sum_{m=1,{\rm odd}}^M
(q_{{\rm D}m}^{L2}+q_{{\rm D}m}^{K2})\sinh\lambda_m^{\rm D}
\coth K\lambda_m^{\rm D}
-2\sum_{m=1,{\rm odd}}^M q_{{\rm D}m}^{L}q_{{\rm D}m}^{K}
{\sinh\lambda_m^{\rm D}\over
\sinh K\lambda_m^{\rm D}}\nonumber\\&&
-{2\over M}
\sum_{m^\prime,m^{\prime\prime}=0}^{(M-1)/2}\left(
q^K_{cm^\prime}q^K_{cm^{\prime\prime}}
+q^L_{cm^\prime}q^L_{cm^{\prime\prime}}\right)
\left(\kappa-1+\cos{2m^\prime\pi\over M}+\cos{2m^{\prime\prime}\pi\over M}\right)
\Bigg]
\eea
Taking $J,L$ large, the factors in front of the integral reduce to
\bea
{{\cal D}_{\cos}^{closed}(J){\cal D}_{\rm odd}^{{\rm D}}(K)
{\cal D}_{\cos}^{closed}(L)\over{\cal D}_{\cos}^{closed}(N+1)}
&\to&\sqrt{N+1\over JL}{e^{K(\sum_{m=1}^{(M-1)/2}\lambda^c_m
-\sum_{m,{\rm odd}}\lambda_m^{\rm D})/2}\over\prod_{m, odd}
\sqrt{1-e^{-2K\lambda_m^{\rm D}}}}
\nonumber\\
&&\hskip-.5in\sqrt{2}
\sqrt{{\prod_{m,odd}\sinh\lambda_m^{\rm D}\over\prod_{m=1}^{(M-1)/2}
\sinh\lambda_m^c}}\left[{{T_0\over2\pi}\prod_{m=1}^{(M-1)/2}
{T_0\over\pi}\sinh{\lambda_m^c}}\right]
\eea
Meanwhile
\bea
iW^{{\rm D}\prime}&\to&-{T_0\over2}\Bigg[
\sum_{m^\prime=1}^{(M-1)/2} (q_{cm}^{K2}+q_{cm}^{L2})\sinh\lambda_m^c
\nonumber\\&&
+\sum_{m^\prime,m^{\prime\prime}=0}^{(M-1)/2}\left(
q^K_{cm^\prime}q^K_{cm^{\prime\prime}}
+q^L_{cm^\prime}q^L_{cm^{\prime\prime}}\right)
\Bigg[{4\over M^2}
\sum_{m=1,{\rm odd}}^MU^{\rm D}_{mm^\prime}U^{\rm D}_{mm^{\prime\prime}}
\sinh\lambda_m^{\rm D}\coth K\lambda_m^{\rm D}\nonumber\\&&
-{2\over M}\left(\kappa-1+\cos{2m^\prime\pi\over M}
+\cos{2m^{\prime\prime}\pi\over M}\right)\Bigg]\nonumber\\
&&-{8\over M^2}
\sum_{m^\prime,m^{\prime\prime}=0}^{(M-1)/2}
q^K_{cm^\prime}q^L_{cm^{\prime\prime}}
\sum_{m=1,{\rm odd}}^MU^{\rm D}_{mm^\prime}U^{\rm D}_{mm^{\prime\prime}}
{\sinh\lambda_m^{\rm D}\over
\sinh K\lambda_m^{\rm D}}\Bigg]
\eea
Just as in the Neumann case it is convenient to absorb the factors in square
brackets into a rescaling of the integration variables
${\bar q}_m=q_{cm}\sqrt{(T_0/\pi)\sinh{\lambda_m^c}}$ for $m=1,\cdots
(M-1)/2$ and ${\bar q}_0=q_0\sqrt{T_0/2\pi}=q_{c0}\sqrt{T_0/\pi}$.
\bea
{\VEV{N+1,\{x^f\}|0,\{x^i\}}^{closed}_{1{\rm loop}}\over
{\cal D}^{\rm closed}(N+1)}&\sim&
\sqrt{2(N+1)\over JL}{e^{K(\sum_{m=1}^{(M-1)/2}\lambda^c_m
-\sum_{m,{\rm odd}}\lambda_m^{\rm D})/2-(K-1)B_0^{\rm D}}\over\prod_{m, odd}
\sqrt{1-e^{-2K\lambda_m^{\rm D}}}}
\nonumber\\
&&\sqrt{{\prod_{m,odd}\sinh\lambda_m^{\rm D}\over\prod_{m=1}^{(M-1)/2}
\sinh\lambda_m^c}}\int d{\bar q}_{m}^Kd{\bar q}_{m}^L
e^{iW^{D\prime}}\\
B_0^{\rm D}&=&B_0-{\lambda_M^{\rm D}\over 2}\nonumber\\
&=&{1\over2}\sinh^{-1} 1-\sinh^{-1}{\sqrt{\kappa\over2}}
={1\over2}\ln{1+\sqrt{2}
\over1+\kappa+\sqrt{\kappa(\kappa+2)}}
\eea
Comparing the prefactors in this formula with the corresponding factors
for the Neumann case we see that the extra factors are
\bea
\sqrt{2(N+1)\over JL}{e^{-K\lambda_M^{\rm D}/2}\sqrt{\sinh\lambda_{M}^{\rm D}}
\over\sqrt{1-e^{-2K\lambda_M^{\rm D}}}}
&\equiv&\sqrt{N+1\over JL}\sqrt{\eta^{K-1}{1-\eta^2\over1-\eta^{2K}}}
\eea
where for brevity we have defined $\eta\equiv1+\kappa-\sqrt{\kappa(2+\kappa)}
\to2-\sqrt{3}\approx0.268$ for $\kappa=1$.
The $\sqrt{(N+1)/JL}$ factor just reflects the fact that the intermediate
open string has its ends fixed in space. In the Neumann case this
factor would instead be $1$. The factor $\eta^{K-1}$
can be absorbed in the boundary counterterm, converting its
zero coupling value back to $B_0$. The expansion in
powers of $\eta$ represents excitations of order $\mathcal{O}(1)$ in
lattice units, which will be suppressed in the continuum physics.

Finally we turn to the matrix determinant that results from the
execution of the Gaussian integration. For this we need to spell
out $W^{D\prime}$ which remains after integrating out the closed string
sine modes. Expressed in terms of the new variables ${\bar q}_{m^\prime}$,
we write
\bea
iW_{\rm D}^{\prime\prime}&\equiv&-\pi\left[\sum_{m^\prime,m^{\prime\prime}}
({\bar q}^K_{m^\prime}{\bar q}^K_{m^{\prime\prime}}
+{\bar q}^L_{m^\prime}{\bar q}^L_{m^{\prime\prime}})A^{\rm D}_{m^\prime
m^{\prime\prime}}+2\sum_{m^\prime,m^{\prime\prime}}
{\bar q}^K_{m^\prime}{\bar q}^L_{m^{\prime\prime}}B^{\rm D}_{m^\prime
m^{\prime\prime}}\right]\\
A^{\rm D}_{00}&=&{2\over M^2}
\sum_{m=1,{\rm odd}}^MU^{\rm D}_{m0}U^{\rm D}_{m0}
\sinh\lambda_m^{\rm D}\coth K\lambda_m^{\rm D}-{1+\kappa\over M}\\
A^{\rm D}_{0m^\prime}=A^{\rm D}_{m^\prime0}&=&{2\over M^2}
\sum_{m=1,{\rm odd}}^MU^{\rm D}_{mm^\prime}U^{\rm D}_{m0}
{\sinh\lambda_m^{\rm D}\coth K\lambda_m^{\rm D}
\over\sqrt{\sinh\lambda^c_{m^\prime}}}
-{\kappa+\cos{2m^\prime\pi/M}\over M\sqrt{\sinh\lambda^c_{m^\prime}}}\\
A^{\rm D}_{m^\prime m^{\prime\prime}}&=&
{\delta_{m^\prime m^{\prime\prime}}\over2}
+{2\over M^2}
\sum_{m=1,{\rm odd}}^MU^{\rm D}_{mm^\prime}U^{\rm D}_{mm^{\prime\prime}}
{\sinh\lambda_m^{\rm D}\coth K\lambda_m^{\rm D}
\over\sqrt{\sinh\lambda^c_{m^\prime}}
\sqrt{\sinh\lambda^c_{m^{\prime\prime}}}}\nonumber\\&&
-{\kappa-1+\cos(2m^\prime\pi/M)
+\cos(2m^{\prime\prime}\pi/M)
\over M\sqrt{\sinh\lambda^c_{m^\prime}}
\sqrt{\sinh\lambda^c_{m^{\prime\prime}}}}\\
B^{\rm D}_{00}&=&-{2\over M^2}
\sum_{m=1,{\rm odd}}^MU^{\rm D}_{m0}U^{\rm D}_{m0}
{\sinh\lambda_m^{\rm D}\over\sinh K\lambda_m^{\rm D}}\\
B^{\rm D}_{0m^\prime}=B^{\rm D}_{m^\prime0}&=&-{2\over M^2}
\sum_{m=1,{\rm odd}}^MU^{\rm D}_{mm^\prime}U^{\rm D}_{m0}
{\sinh\lambda_m^{\rm D}\over\sinh K\lambda_m^{\rm D}
\sqrt{\sinh\lambda^c_{m^\prime}}}\\
B^{\rm D}_{m^\prime m^{\prime\prime}}&=&-{2\over M^2}
\sum_{m=1,{\rm odd}}^MU^{\rm D}_{mm^\prime}U^{\rm D}_{mm^{\prime\prime}}
{\sinh\lambda_m^{\rm D}\over\sinh K\lambda_m^{\rm D}
\sqrt{\sinh\lambda^c_{m^\prime}}\sqrt{\sinh\lambda^c_{m^{\prime\prime}}}}
\eea
For a D$p$-brane there are $D-p-1\to25-p$ coordinates satisfying Dirichlet
boundary conditions. Putting everything together we have for the zero
energy
amplitude for a closed string tachyon scattering off a D$p$-brane:
\bea
-a{\cal M}_{G,closed}&=&
M\sum_{K=2}^\infty\left[
\left({\coth M\sinh^{-1}1\over M\sqrt{2}}\right)^{1/4}
{e^{K\sum_{m=1}^{M-1}
(\lambda^c_m-\lambda^o_m)/2-(K-1)B_0}
\over\prod_{m=1,odd}^{M-1}\sqrt{1-e^{-2K\lambda^o_m}}}\right]^{24}\nonumber\\
&&\left[{\det}^{-1/2}\pmatrix{A&B\cr B&A\cr}\right]^{p-1}
\left[\sqrt{2\pi\over MT_0}\sqrt{{1-\eta^2\over1-\eta^{2K}}}\
{\det}^{-1/2}
\pmatrix{A^{\rm D}&B^{\rm D}\cr B^{\rm D}&A^{\rm D}\cr}\right]^{25-p}\\
\eta&=&1+\kappa-\sqrt{\kappa(\kappa-1)}
\label{dn_shift_closed_groundstate}
\eea
where the scattering amplitude is obtained from the one loop correction to the
two closed string function by stripping off the factor
$\sqrt{MT_0(N+1)/2\pi JL}$ for each Dirichlet dimension,
as explained at the end of
Appendix~\ref{propagators}. The $K=1$ term is not included in the scattering
amplitude since it contributes to the $I$ term of the S-matrix.
\subsection{Graviton Scattering off D-brane}
Let us take the graviton polarizations to lie within the D$p$-brane.
Then in parallel to the derivation of (\ref{shift_graviton}) we
must simply insert the factors
\bea
&&\hskip-.5in2T_0^2\sinh^2\lambda_1^c(\VEV{q_{L,1}^{c1}q_{K,1}^{c1}}^2
+\VEV{q_{L,1}^{s1}q_{K,1}^{s1}}^2)e^{2K\lambda_1^c}
={1\over2}(1+(B_{11}^\prime e^{K\lambda_1^c})^2)
\eea
into the $K$ summand for the closed string tachyon scattering amplitude.

\section{Numerics of D-branes}\label{sec_dbranes_numerics}

In the case of a closed string tachyon scattering off a D$p$-brane,
it is convenient to define the quantity
\be
r_K=\sqrt{\frac{\det(A+B)\det(A-B)}{M\det(A^D+B^D)\det(A^D-B^D)}}\,,
\ee
such that the corresponding zero energy amplitude (\ref{dn_shift_closed_groundstate}) may be rewritten as
\be
-\frac{a{\cal M}_{G,closed}}{M}=
\sum_{K=2}^\infty \delta P^-_K \Bigg(\sqrt{{1-\eta^2\over1-\eta^{2K}}} r_K\Bigg)^{25-p}\,,\label{dn_shift_closed_groundstate_two}
\ee
where $\delta P^-$ is the summand of the tachyon energy shift (\ref{rescaled_DeltaP}). For simplicity we have set $T_0=2\pi$ and also $\kappa=1$, as we have checked that varying its value does not substantially change our results.

\begin{figure}
\includegraphics[width=0.49\textwidth]{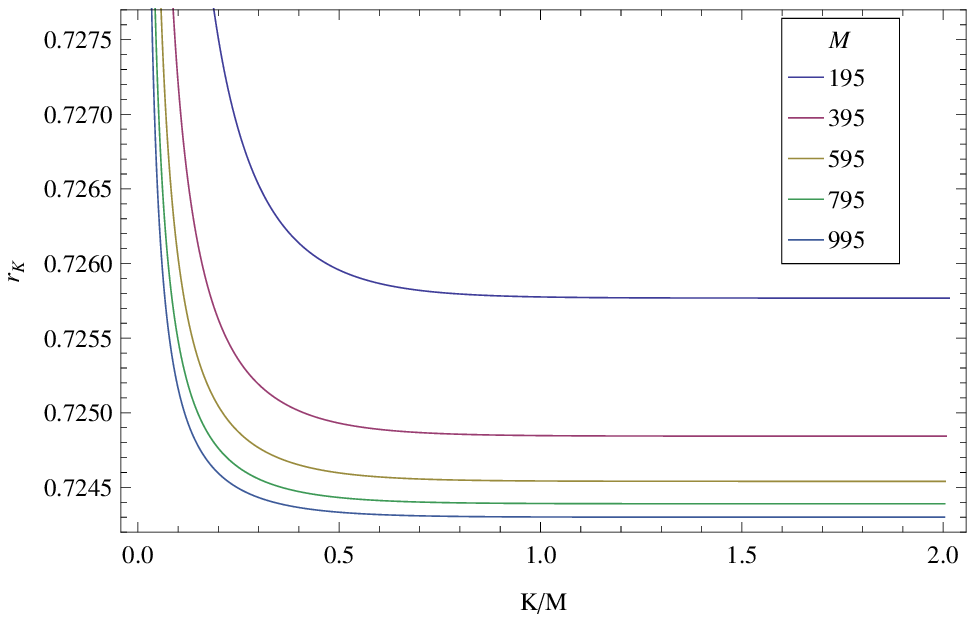}
\includegraphics[width=0.49\textwidth]{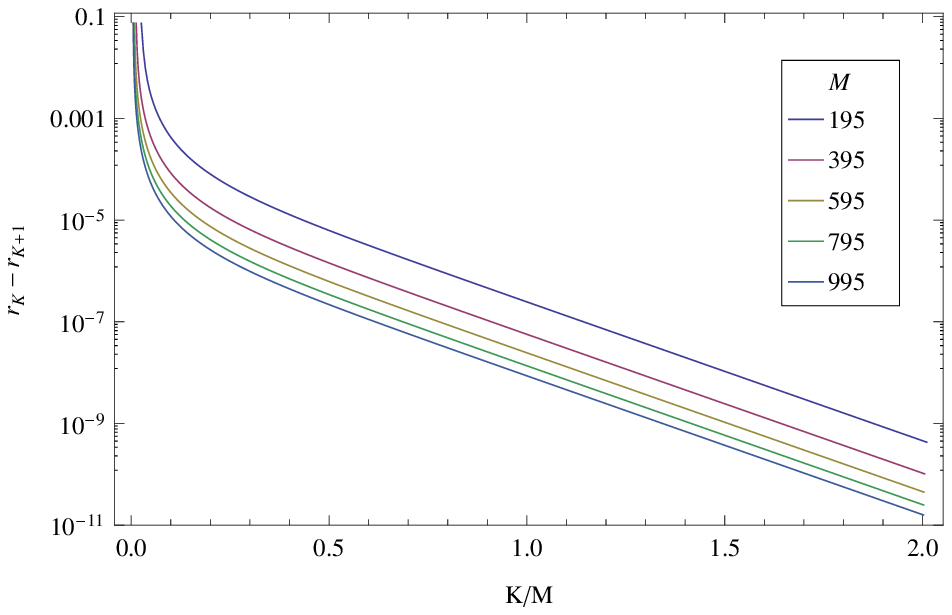}
\caption{\small Linear plot and Log-linear plot of $r_K$ and
its first difference respectively, where each curve corresponds
to fixed $M$ and varying $K$. On the left we see that for $K\gg M$
$r_K$ approaches a value which only depends on $M$, and on the right
in particular that the difference of $r_K$ from this value
is proportional to $e^{-6.28K/M}$ for any $M$ in this regime.}
\label{fig:r_K_large}
\end{figure}
\begin{figure}
\includegraphics[width=0.49\textwidth]{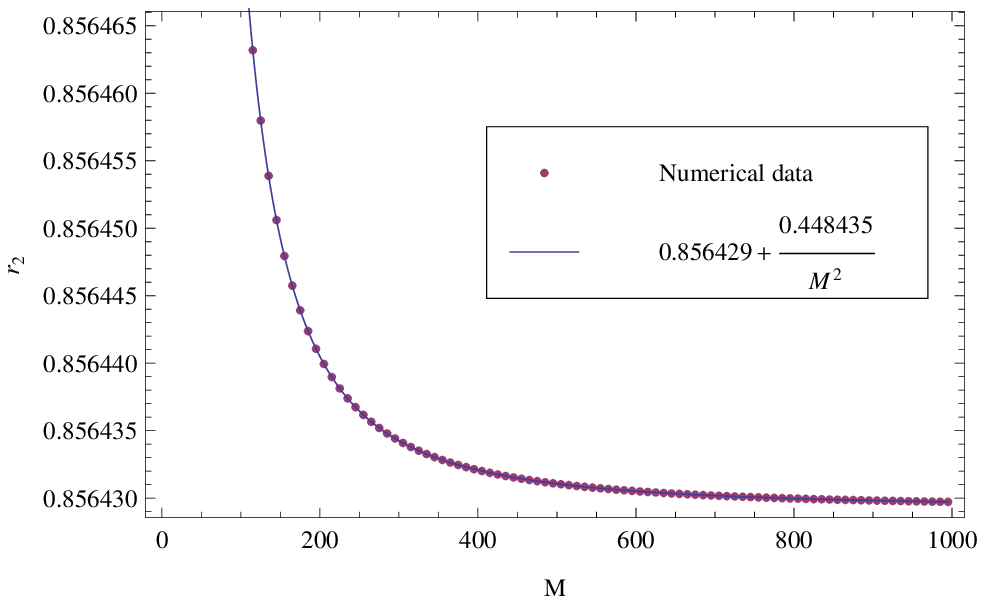}
\includegraphics[width=0.49\textwidth]{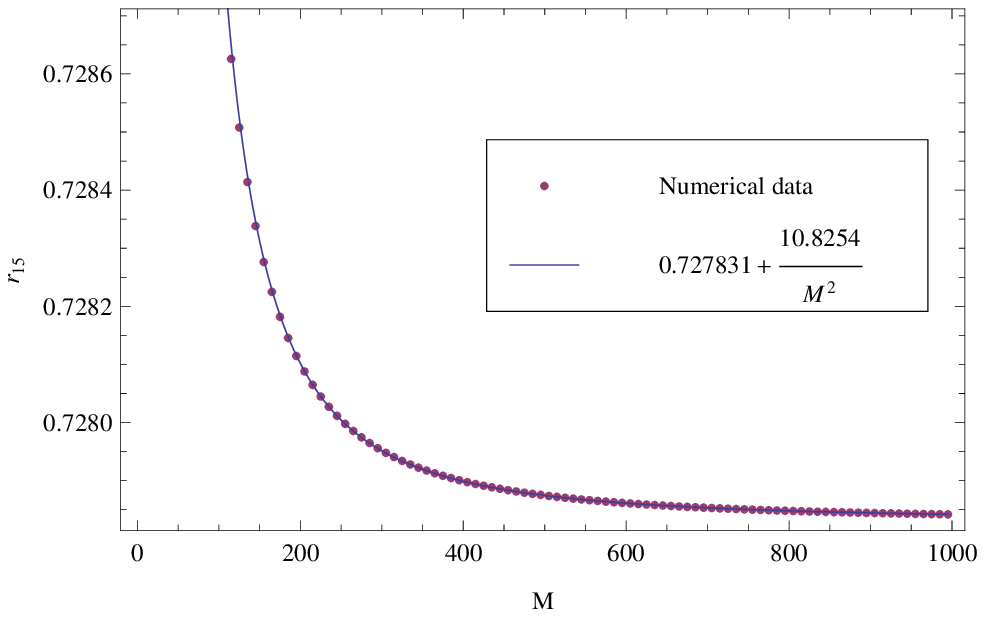}
\caption{\small $r_K$ as a function of $M$ for $K=2, 15$, including fits of the form (\ref{r_K_fit1}).}
\label{fig:r_2-15}
\end{figure}
\begin{figure}
\includegraphics[width=0.49\textwidth]{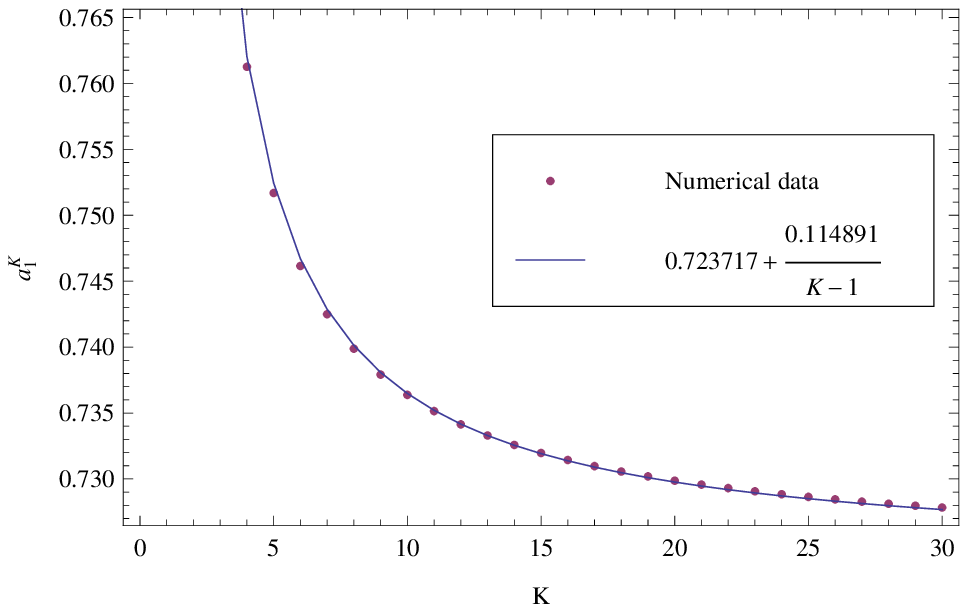}
\includegraphics[width=0.49\textwidth]{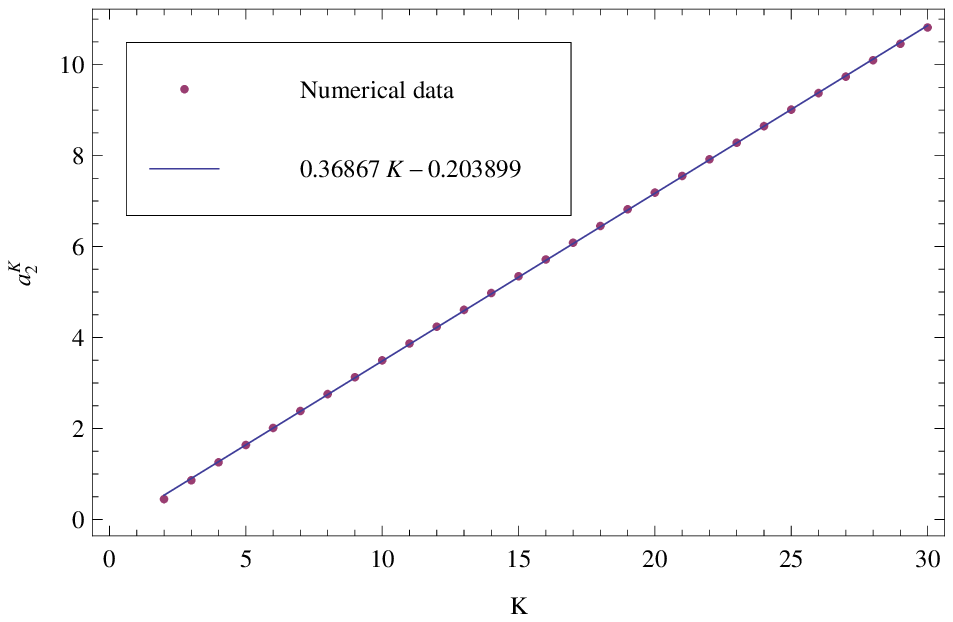}
\caption{\small Values of the fitted coefficients $a_1^K$, $a_2^K$ in (\ref{r_K_fit1}) for $K=1,\ldots,30$ plotted against their estimates (\ref{r_K_fit2a})-(\ref{r_K_fit2}).}
\label{fig:a_1-2}
\end{figure}

Hence the ratio $r_K$ essentially encodes the difference between
Neumann and Dirichlet boundary conditions, and we find that for
$K\gg M$ it falls off exponentially to a value of $\mathcal{O}(1)$
which depends very mildly on $M$, as can be seen
in figure \ref{fig:r_K_large}. For fixed $K\ll M$ we find that the fit
\be\label{r_K_fit1}
r_K=a^K_1+\frac{a^K_2}{M^2}+\mathcal{O}(\frac{1}{M^3})
\ee
matches very well with the data (see figure \ref{fig:r_2-15} for indicative values of $K$), and by further examining the fitted coefficients for different values of $K$, we infer that they roughly vary as
\bea
a_1^K&\approx& 0.724+\frac{0.115}{K-1}\,,\label{r_K_fit2a}\\
a_2^K&\approx& -0.204+0.369K\,,\label{r_K_fit2}
\eea
see also figure \ref{fig:a_1-2}.
By factoring out the constant term in $a_1^K$, which dominates $r_K$, we can infer that the leading dependence of the entire amplitude (\ref{dn_shift_closed_groundstate_two}) on $p$ will be equal to $(0.724\sqrt{1-\eta^2})^{25-p}$ or equivalently $0.697^{25-p}$.


\begin{figure}
\begin{center}
\includegraphics[width=0.83\textwidth]{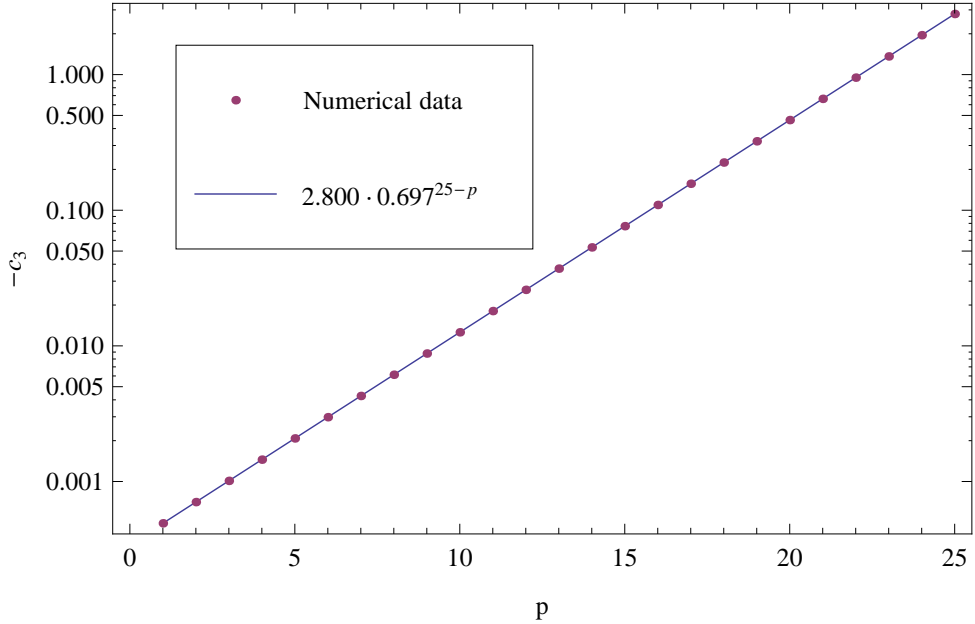}
\caption{\small Log-linear plot of the $\log M/M^2$ coefficient $c_3$ in the large $M$ expansion of the tachyon-D$p$-brane scattering amplitude (\ref{self_energy_fit2}), as a function of $p$.}
\label{fig:dbranesum_c3}
\end{center}
\end{figure}

In fact, it is evident that the only $1/(K M^2)$ term in the summand
(\ref{dn_shift_closed_groundstate_two})
can arise when the aforementioned $p$-dependent factor multiplies the $\mathcal{O}(1/M^2)$ term in $\delta P^-$ (\ref{deltaP_M_expansion}), and given (\ref{harmonicsum}), our analysis constitutes a definite prediction for the relation of the $\log M/M^2$ term of the entire sum in the absence or presence of D-branes. Furthermore, the structure of $r_K$ is such that no $1/K^2$ or $K^{2n-2}/M^{2n}$ terms appear in the summand (\ref{dn_shift_closed_groundstate_two})\footnote{$1/K^2$ would require that $a_1^K$ has at least a linear term in $K$, whereas the $K^l/M^{2n}$ terms that appear always have $l\le n-3$ for any $n$.}, and consequently no $1/M$ terms appear in the sum, so that its expansion to next-to-next-to-leading order in $M$ is expected to be of the the same form as in (\ref{self_energy_fit}).
\begin{figure}
\includegraphics[width=0.49\textwidth]{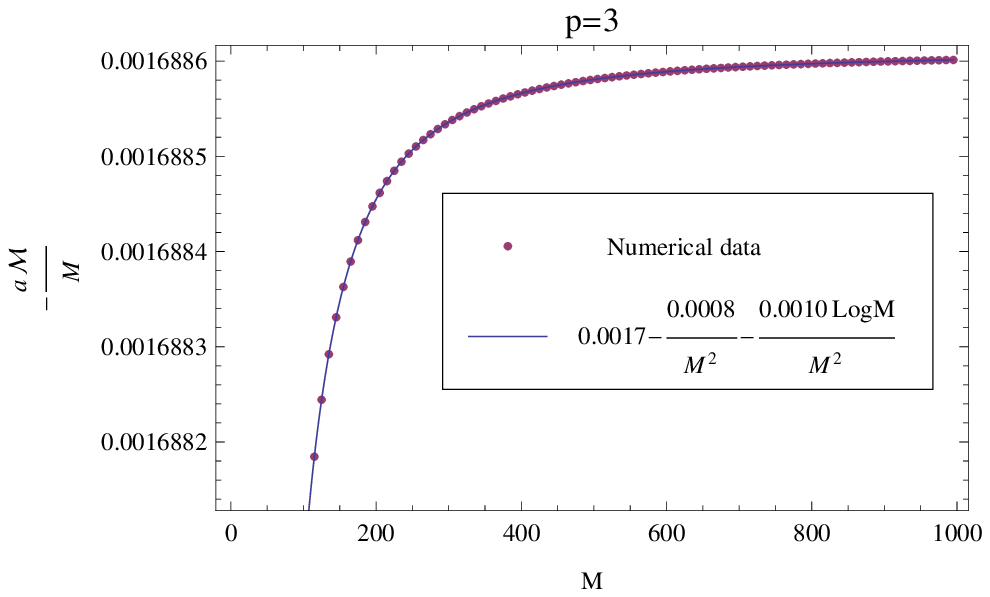}
\includegraphics[width=0.49\textwidth]{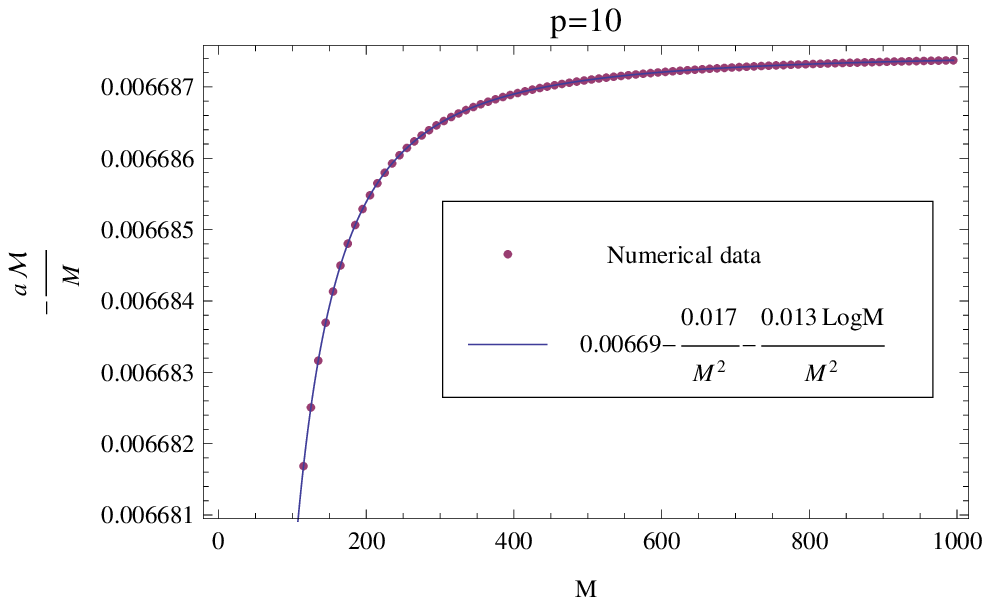}
\includegraphics[width=0.49\textwidth]{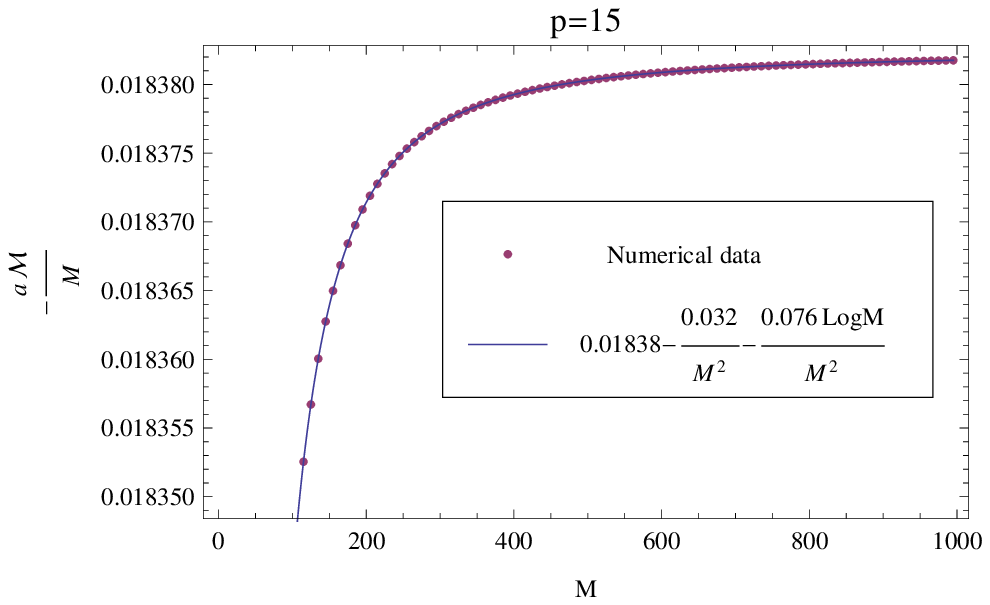}
\includegraphics[width=0.49\textwidth]{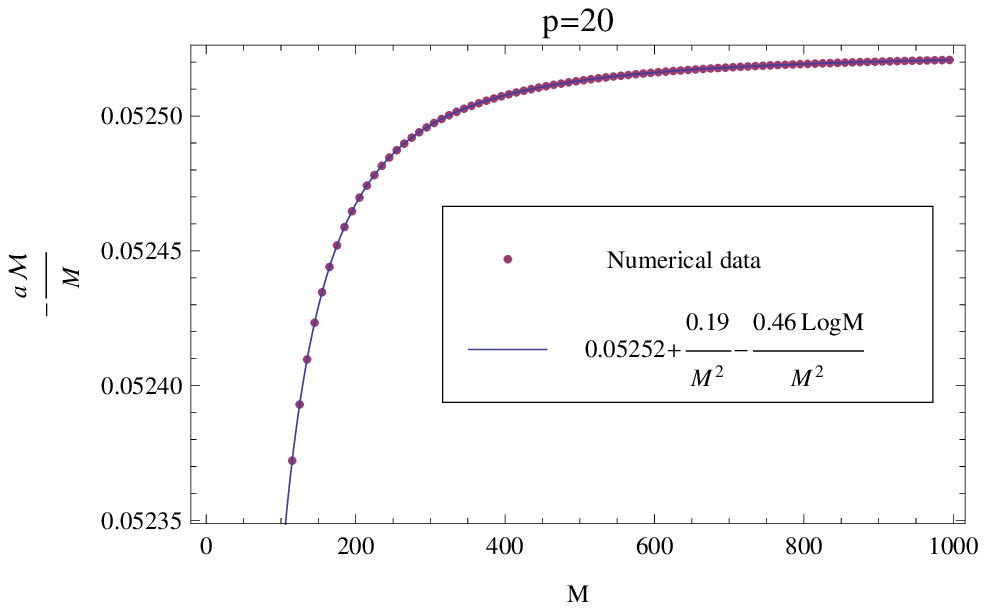}
\caption{\small Rescaled zero energy amplitude for a closed string tachyon scattering off a D$p$-brane for $p=3,10,15,20$ as a function of $M$.}
\label{fig:dbranesum}
\end{figure}
Having gained this insight from the analysis of $r_K$ and the summand, we proceed to evaluate and fit the entire zero energy amplitude (\ref{dn_shift_closed_groundstate_two}), and find that its leading large $M$ behavior is indeed captured by an expansion of the form
\beq\label{self_energy_fit2}
-\frac{a{\cal M}_{G,closed}}{M}=c_1+c_2 \frac{1}{M^2}+c_3\frac{\log M}{M^2}\,.
\eeq
In figure \ref{fig:dbranesum} we compare the fit with the numerics and for sample values of $p$, and give for these cases the values of the $p$-dependent coefficients. It's worth taking a closer look at the dependence of $c_3$ on $p$ in order to see the effect of the presence of D-branes on the undesirable logarithmic divergence, and also verify the prediction for its value based on the aforementioned $r_K$ analysis. Plotting $\log(-c_3)$ as a function of $p\in[1,25]$ (see figure \ref{fig:dbranesum_c3}) we identify a clear linear dependence, which we can fit in order to obtain
\be
c_3=-2.800\cdot (0.697)^{25-p}\,.
\ee
This is precisely the coefficient of the corresponding term in the closed string tachyon energy shift (\ref{self_energy_fit_coefficients}) times the factor we identified below (\ref{r_K_fit2}). The dependence of $c_3$ on $p$ shows that the presence of $D$-branes only softens the divergence in the sense of reducing its coefficient, but cannot remove it completely. Furthermore, these results serve as additional evidence that the existence of the divergence is solely due to the summation of the term in the summand of the closed string tachyon self-energy, which behaves as $1/K$ for small $K$.

Finally, let us conclude by briefly examining what changes when instead of a tachyon we have a graviton whose polarizations lie within the D-brane. We first recall that in the absence of D-branes, the summand for the graviton rescaled energy shift (\ref{grav_summand_fit}) has no $1/M^2$ term, meaning that the graviton remains massless at short distances. On the contrary here we notice that the presence of such a term in $r_K$ (\ref{r_K_fit1}) will carry through to the graviton summand, which can in turn be interpreted as mass generation due to the explicit breaking of Lorentz invariance by the Dirichlet boundary conditions. Although in this particular case the mass is tachyonic, generally having a mechanism for mass generation may be viewed a desirable feature, as we are ultimately interested in using open string theory to probe QCD phenomena. Thus the massless spin-1 states of the former will have to acquire a mass if they are to be put in correspondence with massive gluonic states.

In more detail, if our previous fits (\ref{grav_summand_fit})-(\ref{r_K_fit1}) have captured the $M$-dependence of each factor correctly, we should have
\bea
-\frac{a{\cal M}_{Graviton}}{M}&=&
\sum_{K=2}^\infty [\frac{1}{2}(1+C_G^K)\delta P^-_K ]\Bigg(\sqrt{{1-\eta^2\over1-\eta^{2K}}} r_K\Bigg)^{25-p}\label{dbrane_gravsum}\\
&=&\sum_{K=2}^\infty\Big(\tilde c_1^K+\frac{\tilde c_2^K}{M^4}+\mathcal{O}(\frac{1}{M^5})\Big)\Bigg(a^K_1+\frac{a^K_2}{M^2}+\mathcal{O}(\frac{1}{M^3})\Bigg)^{25-p}\nonumber\\
&=&\sum_{K=2}^\infty\Big[\tilde c_1^K (a^K_1)^{25-p}+(25-p)\tilde c_1^K(a^K_1)^{24-p}a^K_2\frac{1}{M^2}+\mathcal{O}(\frac{1}{M^3})\Big]\,.\label{dn_shift_closed_groundstate_two2}
\eea
\begin{figure}
\begin{center}
\includegraphics[width=0.83\textwidth]{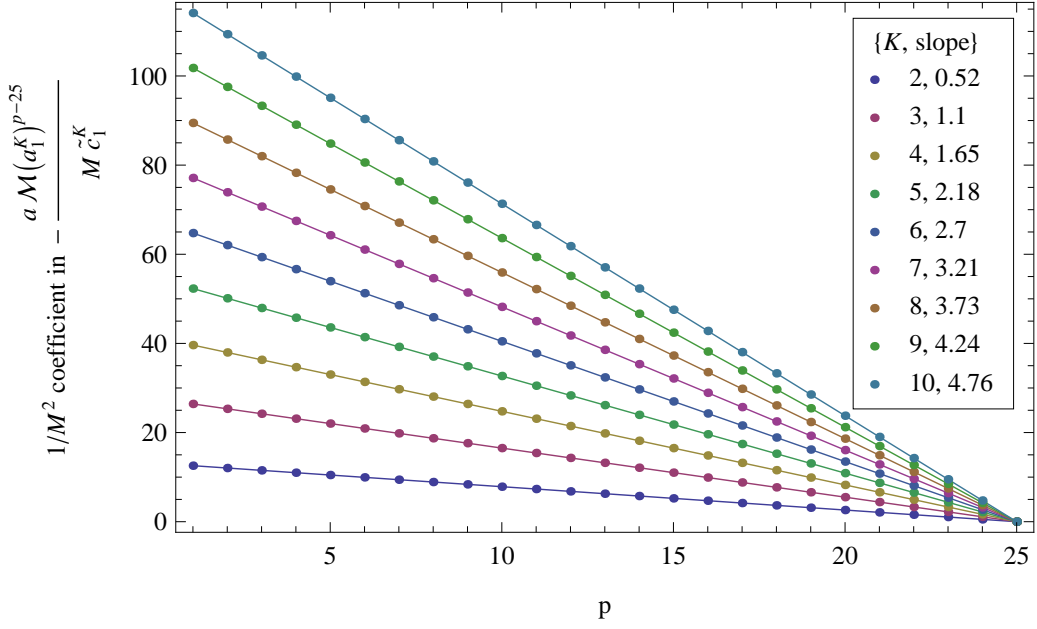}
\caption{Ratio of the first two coefficients in the large $M$ expansion of the summand of the tachyon-D$p$-brane scattering amplitude (\ref{dn_shift_closed_groundstate_two2}), as a function of $p$. Each line represents the $K$-th term in the sum, $K=2,\ldots,10$, and the value of the corresponding slope is also provided in the legend.}
\label{dbrane_grav}
\end{center}
\end{figure}
In particular the ratio of the second to first term should be a linear function of $p$ with slope $a_1^K/a_2^K$ which always becomes zero for $p=25$, as we will now explicitly confirm. We fit the summand in (\ref{dbrane_gravsum}) for $p=1\,\ldots,25$ and $K=2,\ldots,10$ to inverse powers of $M$ up to $\mathcal{O}(1/M^4)$, take the ratio of the coefficients of the first two terms and perform linear regression in terms of $p$ for each value of $K$. As can be seen in figure \ref{dbrane_grav}, the dependence of the ratio of the coefficients on $p$ is clearly linear with the right intercept, and comparing the slope to $a_1^K/a_2^K$ we find that in all cases it differs less than $0.1\%$.

\section{Discussion and Conclusion}\label{conclusion}
In this article we have made a modest beginning to a critical study of the
effectiveness of the worldsheet lattice, introduced in
\cite{gilest}, for implementing a regulated bosonic string loop expansion.
We have limited our analysis to the one-loop corrections to
the closed string two point function in the presence of D-branes,
but we plan to analyze the open string self-energy in a second paper.

A convenient way to summarize our results is to recapitulate
the lightcone lattice interpretation of the divergences in the
well-known covariant expression for the closed string
tachyon self-energy (\ref{secft}):
\bea
-\Delta P^-&=& {C\over2P^+}\int_0^1\left[{dq\over q^3}-2{dq\over q}+qdq\right]
\eea
The first term $dq/q^3$ gives a quadratic divergence,
which in the covariant description
is associated with the closed string tachyon disappearing into the
vacuum. On the lightcone worldsheet lattice, we have seen that
this divergence is interpreted as a contribution to $P^-\sim-\alpha M$,
which can be cancelled by the bulk counterterm proportional to the
area $M(N+1)$ of the worldsheet lattice. The only states that survive
the continuum limit are those with the smallest  value of $\alpha$.
Since only energy differences are physically significant,
this shows that the quadratic divergence is physically inconsequential.

The second term $dq/q$ gives a logarithmic divergence, associated with the
disappearance of a closed string dilaton into the vacuum.
This divergence can be absorbed into a renormalization of the
Regge slope parameter $\alpha^\prime$. Our analysis using the
worldsheet lattice has shown that
this divergence shows up as a contribution to $P^-$ behaving
as $(a M)^{-1}\ln M=(T_0/P^{+})\ln(P^+/aT_0)$. The noncovariance of the
finite residuum was caused by a cutoff on the $K=T/a$ sum
proportional to $M$. If the cutoff were $M$ independent the
noncovariance would disappear. We then noted that such
a cutoff is naturally introduced by adding a small positive
number to the boundary counterterm $B_0\to B_0+\epsilon$.
Then the $\ln M$ is replaced by $\ln(1/\epsilon)$ which can
then be covariantly absorbed in a renormalization of $\alpha^\prime$,
{\it before} the continuum limit $M\to\infty$.

This same $\epsilon$ prescription also prevents the graviton from
gaining a mass in the absence of D-branes. Interestingly, in the
presence of a D-brane the low energy amplitude for a graviton scattering
off the D-brane, which can be thought of as a ``self-energy'' for a graviton
propagating parallel to the D-brane, suggests a nonzero
graviton ``mass'' in spite of the $\epsilon$ prescription. This is a
consistent outcome since Lorentz invariance is broken by the Dirichlet
boundary conditions. This
is important in applying these ideas to large N QCD,
since the closed string is supposed to
model glueballs, all of which should be massive.

At a more fundamental level, we can recognize $\epsilon$ as a
natural and {\it bona fide} physical parameter of the lightcone
worldsheet system: it is simply a measure of the boundary energy
$B\equiv B_0+\epsilon$ associated with the disappearance of a link.
It is also the minimum energy assigned to a free string bit.
The free open string has a Lorentz invariant spectrum for only
one value $B=B_0$ (or $\epsilon=0$), but it makes sense to study
the physics of the system as a function of $B$ (or $\epsilon$).
For sufficiently large $B$,
\bea
B>{2G\over\pi},\qquad \epsilon>{2G\over\pi}-{1\over2}\sinh^{-1}(1)
\approx 0.1424350145,
\eea
$M$ free string bits will have an energy greater than the ground
state energy of a closed string of size $M$. In fact, for $B$
this big, any system of open strings with total bit number $M$
has energy larger than $P^-_{\rm closed, G}$. Thus for $B>2G/\pi$,
the free closed string is stable against decay into any number
of open strings, and our scheme to sum planar diagrams should make
good physical sense.

Dialing the value of $B$ gives us a new tool to analyze the
fate of open string tachyons in the bosonic string on the lightcone worldsheet
lattice. One can attempt to simulate the sum over all patterns
of missing links, as explained in the introduction,
in the closed string propagator for a range of
$B$ values, and then study how the physics changes
as $B$ is gradually decreased.
If the dynamics is favorable, the system should find the true
vacuum with all traces of tachyons removed.
Of course it is possible that the dynamics does not stabilize the
theory lending weight to the prevailing opinion that the
presence of tachyons of the
bosonic string theory is an incurable disease which can only
be cured by replacing
the bosonic string with the superstring.

By introducing a D3-brane and suitable orbifold projections, it is
possible to arrange the gauge boson sector of the open string
to enjoy the same dynamics as gauge field theory in 4 space-time
dimensions. Of course the
open and closed string tachyons are still present, but
they could simply be a symptom that perturbation theory is being
attempted about the ``wrong'' vacuum. If so, analyzing the worldsheet
lattice system as a function of $B$ could provide a way to
find the ``right'' vacuum. It will be an interesting exercise to
apply these techniques to the problem of quark confinement.

\vskip14pt
\noindent\underline{ Acknowledgments}:
This research was supported in part by the Department of
Energy under Grant No. DE-FG02-97ER-41029.
\vskip18pt
\appendix

\section{Normal Modes}
\label{normalmodes}
A string with $P^+=MaT_0$ is described at a fixed time by
$M$ coordinates $x_i$ or $y_i$, $i=1,\cdots M$. In this article we require
several normal mode decompositions depending on the boundary conditions.
\vskip12pt
\noindent Neumann Open String
\bea
x_i&=&{1\over\sqrt{M}}q_{0}+\sqrt{2\over M}\sum_{m=1}^{M-1}q_{om}
\cos{m\pi(i-1/2)\over M}\\
q_0&=&\sqrt{1\over M}\sum_{i=1}^M x_i,\qquad\quad q_{om}=\sqrt{2\over M}\sum_i x_i\cos{m\pi(i-1/2)\over M}
\label{normalNopen}
\eea
\vskip12pt
\noindent Closed String (Neumann)
\bea
&&\hskip-1inM~{\rm odd}:\nonumber\\
x_i&=&{1\over\sqrt{M}}q_{0}+\sqrt{2\over M}\sum_{m=1}^{(M-1)/2}\left[q_{cm}
\cos{2m\pi(i-1/2)\over M}+q_{sm}\sin{2m\pi(i-1/2)\over M}\right]\\
&&\hskip-1inM~{\rm even}:\nonumber\\
x_i&=&{1\over\sqrt{M}}(q_{0}+q_{sM/2}(-)^i)
\nonumber\\&&
+\sqrt{2\over M}\sum_{m=1}^{M/2-1}\left[q_{cm}
\cos{2m\pi(i-1/2)\over M}+q_{sm}\sin{2m\pi(i-1/2)\over M}\right]\\
q_{cm}&=&\sqrt{2\over M}\sum_i x_i
\cos{2m\pi(i-1/2)\over M},\qquad q_{sm}=\sqrt{2\over M}\sum_i x_i
\sin{2m\pi(i-1/2)\over M}\\
q_{sM/2}&=&\sqrt{1\over M}\sum_{i=1}^M (-)^ix_i,
\qquad {\rm for}\quad M\quad{\rm even},\qquad
q_0=\sqrt{1\over M}\sum_{i=1}^M x_i
\label{normalNclosed}
\eea
\vskip12pt
\noindent Dirichlet Open String
\bea
y_k&=&\sqrt{2\over M}\sum_{m=1}^{M-1}q_{{\rm D}m}
\sin{m\pi k\over M}\quad{\rm for}\quad k=1,\cdots,M-1,\qquad
y_M=q_{{\rm D}M}\\
q_{{\rm D}m}&=&\sqrt{2\over M}\sum_{k=1}^{M-1}y_k\sin{m\pi k\over M},\quad
0<m<M,\qquad
q_{{\rm D}M}=y_M
\label{normalDopen}
\eea
\vskip12pt
\noindent Closed String (Dirichlet)
\bea
&&\hskip-1inM~{\rm odd}:\nonumber\\
y_i&=&{1\over\sqrt{M}}q_{0}+\sqrt{2\over M}\sum_{m=1}^{(M-1)/2}\left[q_{cm}
\cos{2m\pi i\over M}+q_{sm}\sin{2m\pi i\over M}\right]\\
&&\hskip-1inM~{\rm even}:\nonumber\\
y_i&=&{1\over\sqrt{M}}(q_{0}+q_{cM/2}(-)^i)
+\sqrt{2\over M}\sum_{m=1}^{M/2-1}\left[q_{cm}
\cos{2m\pi i\over M}+q_{sm}\sin{2m\pi i\over M}\right]\\
q_{cm}&=&\sqrt{2\over M}\sum_i y_i
\cos{2m\pi i\over M},\qquad q_{sm}=\sqrt{2\over M}\sum_i y_i
\sin{2m\pi i\over M}\\
q_0&=&\sqrt{1\over M}\sum_{i=1}^M y_i,
\qquad\qquad q_{cM/2}=\sqrt{1\over M}\sum_{i=1}^M (-)^iy_i,
\quad({\rm for}~M~{\rm even})
\label{normalDclosed}
\eea
\section{Determinants}
\label{determinants}
Define the normal mode frequencies of a one dimensional harmonic chain
\bea
\alpha_n&\equiv&4\sin^2{n\pi\over2(N+1)},\qquad n=1,2,\ldots,N\\
\beta_n&\equiv&4\sin^2{m\pi\over2M},\qquad m=0,1,\ldots,M-1\\
\gamma_k&\equiv&4\sin^2{(k+1/2)\pi\over2K+1},\qquad k=0,1,\ldots,K-1\\
\delta_n&\equiv&4\sin^2{m\pi\over M},\qquad m=0,1,\ldots,M-1
\eea
where $\alpha,\beta,\gamma,\delta$ are the modes of a Dirichlet-Dirichlet,
Neumann-Neumann, Dirichlet-Neumann, closed chain respectively.
Then we are interested in the following determinants: (see, for example
\cite{thorndets}:
\bea
D_{\rm DDDD}&=&\prod_{n=1}^N\prod_{m=1}^{M-1}(\alpha_n+\beta_m)^{-1/2},\qquad
D_{\rm DNDN}\ =\ \prod_{n=1}^N\prod_{m=0}^{M-1}(\alpha_n+\beta_m)^{-1/2}\nonumber\\
D_{\rm DDDN}&=&\prod_{m=1}^{M-1}
\prod_{k=0}^{K-1}(\beta_m+\gamma_k)^{-1/2},\qquad
D_{\rm DNNN}\ = \prod_{m=0}^{M-1}
\prod_{k=0}^{K-1}(\beta_m+\gamma_k)^{-1/2}\\
D_{\rm DD\ ring}&=&\prod_{n=1}^{N}
\prod_{m=0}^{M-1}(\alpha_n+\delta_m)^{-1/2},\qquad
D_{\rm ND\ ring}\ =\ \prod_{n=0}^{N-1}
\prod_{m=0}^{M-1}(\gamma_n+\delta_m)^{-1/2}\nonumber
\eea
where the subscripts denote Dirichlet (D) or Neumann (N) boundary
conditions on each of the four edges of the rectangle.
The cylinder determinant with Neumann boundary conditions must
be defined to exclude the overall zero mode:
\bea
D_{\rm NN\ ring}&\equiv&\sqrt{MN}\prod_{(n,m)\neq(0,0)}
(\beta_n+\delta_m)^{-1/2}=\sqrt{MN}\prod_{m=1}^{M-1}\delta_m^{-1/2}
D_{\rm DD\ ring}\nonumber\\
&=&\sqrt{N\over M}D_{\rm DD\ ring}
\eea
The following product identities can be easily derived:
\bea
\prod_{n=1}^N(\alpha_n-z)&=&{\sin(N+1)\kappa\over\sin\kappa},
\qquad\prod_{k=0}^{K-1}(\gamma_k-z)={\cos[(2K+1)\kappa/2]
\over\cos[\kappa/2]}\\
\prod_{m=1}^{M-1}(\delta_m-z)&=&{\sin^2(M\kappa/2)\over
\sin^2(\kappa/2)},\qquad\prod_{m=1}^{M-1}(\beta_m-z)
={\sin M\kappa\over\sin\kappa}
\label{productidentities}
\eea
where $z$ and $\kappa$ are related by $z=4\sin^2[\kappa/2]$.
Applying these identities at $z=0,\kappa=0$ shows immediately
that $D_{\rm DNDN}=D_{\rm DDDD}/\sqrt{N+1}$ and $D_{\rm DNNN}=D_{\rm DDDN}$.

For convenience we collect here the expressions for the open and closed string
propagators at vanishing initial and final coordinates \cite{gilest}.
These quantities are the determinants just discussed in their
various guises. For example ${\cal D}^{\rm open}$ is simply related to
$D_{\rm DNDN}$, with one of the products performed using the identities
(\ref{productidentities}).
\bea
{\cal D}^{\rm open}(N+1)&=&{1\over\sqrt{N+1}}\left({T_0\over2\pi}\right)^{M/2}
\prod_{m=1}^{M-1}\left[{\sinh(N+1)\lambda^o_m
\over\sinh\lambda^o_m}\right]^{-1/2}\equiv{\cal D}_{\rm even}^{\rm open}
{\cal D}_{\rm odd}^{open}\\
{\cal D}_{\rm odd}^{open}(N+1)&=&
\left({T_0\over2\pi}\right)^{(M-1)/4}
\prod_{m=1,{\rm odd}}^{M-1}\left[{\sinh(N+1)\lambda^o_m
\over\sinh\lambda^o_m}\right]^{-1/2}\\
{\cal D}^{closed}(N+1)&=&{1\over\sqrt{N+1}}\left({T_0\over2\pi}\right)^{M/2}
\prod_{m=1}^{M-1}\left[{\sinh(N+1)\lambda^c_m
\over\sinh\lambda^c_m}\right]^{-1/2}\equiv{\cal D}^{\rm closed}_{\cos}
{\cal D}^{\rm closed}_{\sin}\\
{\cal D}^{\rm closed}_{\sin}(N+1)&=&\left({T_0\over2\pi}\right)^{(M-1)/4}
\prod_{m=1}^{(M-1)/2}\left[{\sinh(N+1)\lambda^c_m
\over\sinh\lambda^c_m}\right]^{-1/2}
\eea
where, for simplicity,
we have written these formulas assuming $M$ is odd. If $M$ were even,
appropriate adjustments to the limits of the products must be made.
\section{Propagators}
\label{propagators}
\subsection{Neumann Open String Propagator}
\bea
\VEV{N+1,\{x^f\}|0,\{x^i\}}^{open}&=&{\cal D}^{open}(N+1)e^{iW_{open}}\\
iW_{open}&=&-{T_0\over2}\bigg[{(q_{0,f}-q_{0,i})^2\over N+1}\nonumber\\
&&\hskip-1in+\sum_{m=1}^{M-1}\sinh\lambda^o_m
\left((q_{m,i}^2+q_{m,f}^2)\coth(N+1)\lambda^o_m
-2{q_{m,i}q_{m,f}\over\sinh(N+1)\lambda^o_m}\right)\bigg]\\
\lambda^o_m&=&2\sinh^{-1}\left(\sin{m\pi\over2M}\right)
\eea
Where the $q_m$'s are the normal mode coordinates for the $x$'s.
The right side is the result of doing the integrations over
all the $x_i^j$ with $i=1,\cdots, M$ and $j=1,\cdots N$. The
propagator spans $N+1$ time steps and this result corresponds to
assigning half the potential energy $T_0\sum_{i=1}^{M-1}(x_{i+1}^j
-x_i^j)^2/2$ to time $j=0$ and half to $j=N+1$.
\subsection{Dirichlet Open String Propagator}
The Dirichlet open string propagator over a time of $K=N+1$ steps is
evaluated to be
\bea
\VEV{N+1,\{q^f\}|0,\{q^i\}}^{{\rm D}}&=&{\cal D}^{\rm D}(N+1)e^{iW^{\rm D}}
\eea
where
\bea
iW^{\rm D}&=&-{T_0\over2}\left[\sum_{m=1}^{M}\left((q_{{\rm D}m}^{f2}+q_{{\rm D}m}^{i2})
\sinh\lambda_m^{\rm D}\coth K\lambda_m^{\rm D}
-2q_{{\rm D}m}^{f}q_{{\rm D}m}^{i}{\sinh\lambda_m^{\rm D}\over
\sinh K\lambda_m^{\rm D}}\right)\right]\\
{\cal D}^{\rm D}(N+1)&=&\left({T_0\over2\pi}\right)^{M/2}
\prod_{m=1}^{M}\left[{\sinh(N+1)\lambda^{\rm D}_m
\over\sinh\lambda^{\rm D}_m}\right]^{-1/2}\\
\lambda^{\rm D}_{M}&=&2\sinh^{-1}{\sqrt{\kappa\over2}},
\qquad \lambda^{\rm D}_{m}
=\lambda^o_m=2\sinh^{-1}\sin{m\pi\over2M},\quad m=1,\cdots M-1
\eea
We recall that the above expressions give the the result of integrating over
all the variables $y^j_i$, for $j=1,\cdots,N$, with half the potential
energy assigned to $j=0,N+1$, which is consistent with the
closure requirement.

\subsection{Closed String Propagator}
\bea
\VEV{N+1,\{x^f\}|0,\{x^i\}}^{closed}&=&{\cal D}^{closed}(N+1)e^{iW_{closed}}\\
iW_{closed}&=&-{T_0\over2}\bigg[{(q_{0,f}-q_{0,i})^2\over N+1}\nonumber\\
&&\hskip-1in+\sum_{m=1}^{M-1}\sinh\lambda^c_m
\left((q_{m,i}^2+q_{m,f}^2)\coth(N+1)\lambda^c_m
-2{q_{m,i}q_{m,f}\over\sinh(N+1)\lambda^c_m}\right)\bigg]\\
\lambda^c_m&=&2\sinh^{-1}\left(\sin{m\pi\over M}\right)
\eea
Where the $q_m$'s are the normal mode coordinates for the $x$'s.
When we divide the closed string normal modes into sine and cosine
modes, we arbitrarily call the $m>M/2$ modes sine modes and the
$m<M/2$ modes cosine modes. When $M$ is even, the $M/2$ mode
is not doubled.
The right side is the result of doing the integrations over
all the $x_i^j$ with $i=1,\cdots, M$ and $j=1,\cdots N$. The
propagator spans $N+1$ time steps and this result corresponds to
assigning half the potential energy $T_0\sum_{i=1}^{M}(x_{i+1}^j
-x_i^j)^2/2$ to time $j=0$ and half to $j=N+1$. In sums like these
it is understood that $x_{M+1}^j\equiv x_1^j$. Whenever we concatenate
at a time $j$ propagators with different numbers of missing links, we will
understand that we {\it add} terms $T_0(\Delta x)^2/4$ in the exponent
so that the
potential assigned to time $j$ is that of the system with the least number
of missing links. For example, the concatenation of an open string
propagator with a closed string propagator entails the addition
of $T_0(x_{M}^j-x_1^j)^2/4$ to the exponent.

Finally, we resolve the zero mode dependence of the propagators
in momentum space
\bea
\int{dp\over2\pi}e^{-p^2T/2P^+}e^{i(x^{CM}_f-x^{CM}_i)p}&=&
\sqrt{P^+\over2\pi T}
e^{-P^+(x^{CM}_f-x^{CM}_i)^2/2T}\nonumber\\
&=&\sqrt{ MT_0\over2\pi (N+1)}
e^{-MT_0(x^{CM}_f-x^{CM}_i)^2/2(N+1)}\nonumber\\
\int{dp\over2\pi}e^{-p^2T/2P^+}e^{i(x^{CM}_f-x^{CM}_i)p}
&=&\sqrt{MT_0\over2\pi(N+1)}
e^{-T_0(q_{0,f}-q_{0,i})^2/2(N+1)}\; ,
\eea
where $x^{CM}\equiv\sum_k x_k/M=q_0/\sqrt{M}$ is the center of
mass coordinate.
From this we see that in extracting eigenstate amplitudes from
propagators defined with Dirichlet conditions on initial and
final states, we must not include the factors $\sqrt{MT_0/2\pi(N+1)}$ .
\section{Overlap Formulas}
\label{overlap}
\vskip12pt
\noindent{Neumann Open-Closed}
\bea
q_{om}&=&\cases{q_{cm/2} &$m$ {\rm even}\cr&\cr
{2\over M}\sum_{m^\prime=1}^{(M-1)/2}q_{sm^\prime}U_{mm^\prime} &
$m$ {\rm odd}\cr}\\
U_{mm^\prime}&=&{\sin(m^\prime\pi/M)\cos(m\pi/2M)\over
\sin^2(m^\prime\pi/M)-\sin^2(m\pi/2M)}
\eea
\vskip12pt
\noindent{Dirichlet Open-Closed}
\bea
q_{{\rm D}M}&=&{1\over\sqrt{M}}q_{0}
+\sqrt{2\over M}\sum_{m^\prime=1}^{(M-1)/2}q_{cm^\prime}\equiv
\sqrt{2\over M}\sum_{m^\prime=0}^{(M-1)/2}q_{cm^\prime}\\
q_{{\rm D}m}&=&q_{sm/2},\qquad {\rm for}\quad m~{\rm even}\\
q_{{\rm D}m}&=&{\sqrt{2}\over{M}}q_{0}\sum_{k=1}^{M-1}\sin{m\pi k\over M}
+{2\over M}\sum_{m^\prime=1}^{(M-1)/2}q_{cm^\prime}
\sum_{k=1}^{M-1}\sin{m\pi k\over M}\cos{2m^\prime\pi k\over M}\nonumber\\
&\equiv&{2\over M}\sum_{m^\prime=0}^{(M-1)/2}q_{cm^\prime}
\sum_{k=1}^{M-1}\sin{m\pi k\over M}\cos{2m^\prime\pi k\over M}
\eea
where, for convenience, we have defined $q_{c0}\equiv q_0/\sqrt{2}$.
The sum over $k$ is easily done
\bea
\sum_{k=1}^{M-1}\sin{m\pi k\over M}\cos{2m^\prime\pi k\over M}
&=&{\delta_{m\ {\rm odd}}\over2}{\sin(m\pi/M)\over\sin^2(m\pi/2M)
-\sin^2(m^\prime\pi/M)}\equiv\delta_{m\ {\rm odd}}U^{\rm D}_{mm^\prime}
\eea
for odd $m<M$. We can unify the treatment of the $m=M$ mode by defining
$U^{\rm D}_{Mm^\prime}=\sqrt{M/2}$ for $m^\prime=0,1,\cdots,(M-1)/2$:
\bea
q_{{\rm D}m}&=&\cases{{\displaystyle q_{sm/2}}& for $m$ even\cr
&\cr
{\displaystyle {2\over M}\sum_{m^\prime=0}^{(M-1)/2}q_{cm^\prime}
U^{\rm D}_{mm^\prime}}& for $m$ odd\cr}
\eea
\vskip12pt
\noindent{Open-2 Open}
\bea
q_0^{(1)}&=&\sqrt{M_1\over M}\ q_0+\sqrt{2\over MM_1}\sum_{m^\prime=1}^{M-1}
 q_{m^\prime}U^{(1)}_{m^\prime 0},\qquad q_m^{(1)}={2\over\sqrt{MM_1}}\sum_{m^\prime=1}^{M-1}
q_{m^\prime}U^{(1)}_{m^\prime m}\\
q_0^{(2)}&=&\sqrt{M_2\over M}\ q_0+\sqrt{2\over MM_2}\sum_{m^\prime=1}^{M-1}
 q_{m^\prime}U^{(2)}_{m^\prime 0},\qquad q_m^{(2)}={2\over\sqrt{MM_2}}\sum_{m^\prime=1}^{M-1}
q_{m^\prime}U^{(2)}_{m^\prime m}\\
U^{(1)}_{m^\prime m}&=&\sum_{i=1}^{M_1}\cos{m^\prime\pi\over M}\left(
i-{1\over2}\right)\cos{m\pi\over M_1}\left(
i-{1\over2}\right)\nonumber\\
&=&{(-)^m\over2}
{\sin(m^\prime\pi M_1/M)\sin(m^\prime\pi/2M)\cos(m\pi/2M_1)\over
\sin^2(m^\prime\pi/2M)-\sin^2(m\pi/2M_1)}\\
U^{(2)}_{m^\prime m}&=&\sum_{i=1+M_1}^{M}\cos{m^\prime\pi\over M}\left(
i-{1\over2}\right)\cos{m\pi\over M_2}\left(
i-M_1-{1\over2}\right)\nonumber\\
&=&-{1\over2}
{\sin(m^\prime\pi M_1/M)\sin(m^\prime\pi/2M)\cos(m\pi/2M_2)\over
\sin^2(m^\prime\pi/2M)-\sin^2(m\pi/2M_2)}
\eea
and we note the identity $q_0^{(1)}\sqrt{M_1}+q_0^{(2)}\sqrt{M_2}
=q_0\sqrt{M}$, as expected from the fact that $q_0/\sqrt{M}$ is the center
of momentum of the open string.

We can also express the $q$'s in terms of the $q^{(1)},q^{(2)}$'s:
\bea
q_0&=&q_0^{(1)}\sqrt{M_1\over M}+q_0^{(2)}\sqrt{M_2\over M}\\
q_{m^\prime}
&=&\sqrt{2\over MM_1}\left(q_0^{(1)}U^{(1)}_{m^\prime0}
+\sqrt{2}\sum_{m=1}^{M_1-1}q_m^{(1)}U^{(1)}_{m^\prime m}\right)
\nonumber\\&&\qquad
+\sqrt{2\over MM_2}\left(q_0^{(2)}U^{(2)}_{m^\prime0}+\sqrt{2}
\sum_{m=1}^{M_2-1}q_m^{(2)}U^{(2)}_{m^\prime m}\right)
\eea

\section{Robustness of numerical results}
\label{robust}
\begin{figure}
\includegraphics[width=0.83\textwidth]{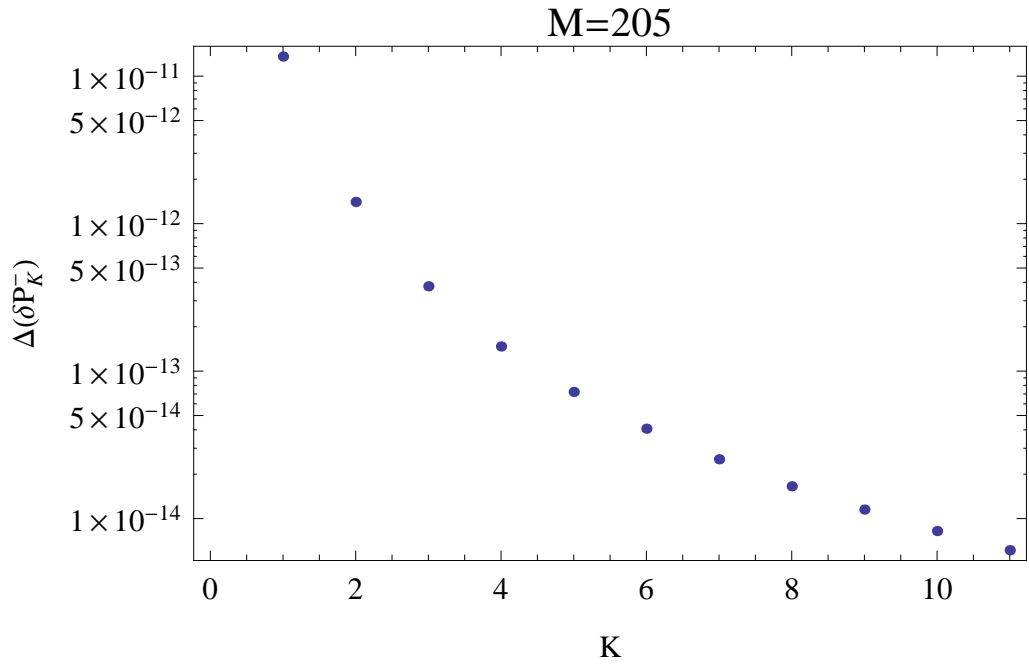}
\caption{\small Log-linear plot of the change in accuracy $\Delta(\delta P^-_K)$ as a function of $K$ for $M=205$.}
\label{fig:error_K}
\end{figure}
In this appendix, we perform additional tests which verify the correctness of our numerical results, within the stated accuracy. In particular, we recalculate $\delta P^-_K$ with more significant digits\footnote{In particular, we now keep 20 significant digits compared to the 16 we had initially.} for indicative values of $M$ and $K$, and examine the improvement in its accuracy.

We will use the difference in $\delta P^-_K$ between the calculations with different number of significant digits, $\Delta(\delta P^-_K)$, as a measure of our numeric error. We first notice that $\Delta(\delta P^-_K)$ decreases almost exponentially when $K$ increases for fixed $M$, as can be seen in the example of figure \ref{fig:error_K}.
\begin{figure}
\includegraphics[width=0.83\textwidth]{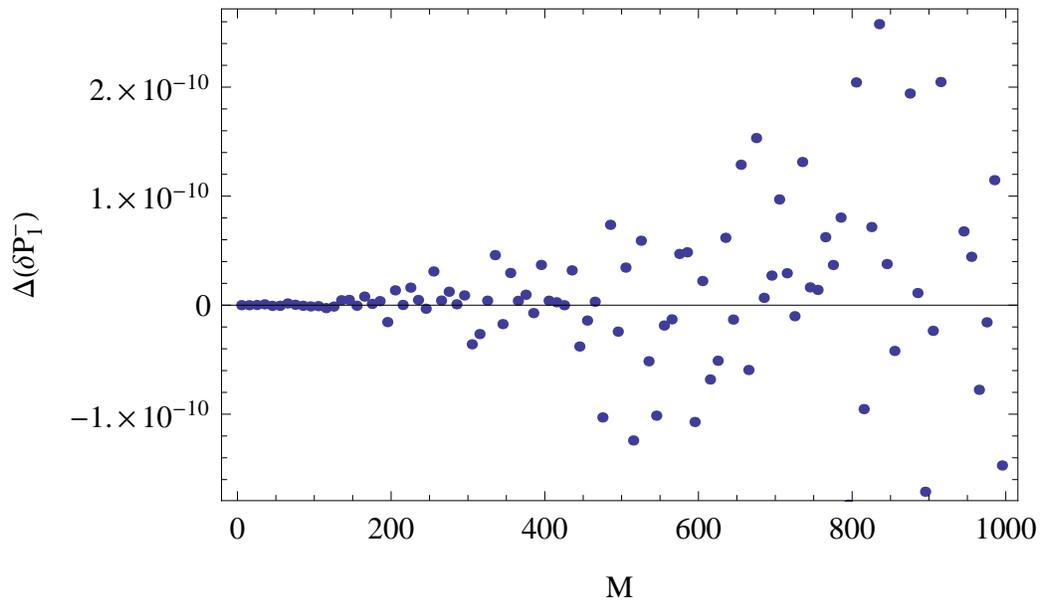}
\caption{\small Plot of the leading change in accuracy $\Delta(\delta P^-_1)$ as a function of $M$.}
\label{fig:error_M}
\end{figure}

The decrease is so rapid that we can clearly consider the error of the $K=1$ term as the error of the entire sum, giving the rescaled ground state energy shift $\Delta P^-_{G,closed}$ in (\ref{rescaled_DeltaP}). Fortunately, in this case we know the exact value $\delta P^-_1=1$ for any $M$, allowing us to also obtain the exact deviation of our numerical results from it, which we present in figure \ref{fig:error_M}. We first notice that the deviations are centered around zero, implying that no systematic error that offsets the value of $\delta P^-_1$ is present. Furthermore, it is evident that the deviation increases with $M$, and for the range $M\in[195,995]$ we have based our fits on, it lies between $10^{-11}-10^{-10}$.

Let us now compare this with the error in the fit for $\Delta P^-_{G,closed}$ (\ref{self_energy_fit}) as a result of the uncertainty in the coefficients (\ref{self_energy_fit_coefficients}),
\beq
\Delta c_1\sim 10^{-9}\,,\quad \Delta \Big(\frac{c_2}{M^2}\Big)\sim 10^{-8}\,,\quad \Delta\Big(\frac{c_3 \log M}{M^2}\Big)\sim 10^{-8}\,\textrm{ for }M\sim 10^{3}\,,
\eeq
where we estimated the smallest possible contribution of the last two terms by replacing $M$ with roughly the largest value we used in our numerical analysis. Clearly these uncertainties are at least one order of magnitude larger than the errors due to our choice for the number of significant digits, so the effect of the latter on the determined values for the coefficients $c_i$ will be negligible. This successfully completes the investigation of the robustness of our numerical results.

\end{document}